\documentclass[10pt]{article}

\usepackage{framed,multirow}
\usepackage{amssymb}
\usepackage{latexsym}
\usepackage{epsfig}
\usepackage{graphicx}
\usepackage{amsmath}
\usepackage{rotating}
\usepackage{listings} 
\usepackage[utf8]{inputenc}
\usepackage[english]{babel}
\usepackage{verbatim}
\usepackage[boxed, longend]{algorithm2e}
\usepackage{color}
\usepackage{dsfont}
\usepackage{authblk}
\usepackage[a4paper, margin=1.0in]{geometry}

\usepackage{newfloat}
\DeclareFloatingEnvironment[
  fileext = los ,
  listname = {List of Schemes} ,
  name = Scheme
]{scheme}

\newcommand{\be}{\begin{eqnarray}}
\newcommand{\ee}{\end{eqnarray}}
\newcommand{\ba}{\begin{array}}
\newcommand{\ea}{\end{array}}
\newcommand{\bmn}{\begin{minipage}}
\newcommand{\emn}{\end{minipage}}
\newcommand{\bv}{\begin{verbatim}}
\newcommand{\ev}{\end{verbatim}}
\newcommand{\bt}{\begin{tabular}}
\newcommand{\et}{\end{tabular}}
\newcommand{\bi}{\begin{itemize}}
\newcommand{\ei}{\end{itemize}}
\newcommand{\btab}{\begin{table}}
\newcommand{\etab}{\end{table}}
\newcommand{\btabs}{\begin{table*}}
\newcommand{\etabs}{\end{table*}}
\newcommand{\btabsw}{\begin{sidewaystable}}
\newcommand{\etabsw}{\end{sidewaystable}}
\newcommand{\bfig}{\begin{figure}}
\newcommand{\efig}{\end{figure}}
\newcommand{\bfigs}{\begin{figure*}}
\newcommand{\efigs}{\end{figure*}}
\newcommand{\bc}{\begin{center}}
\newcommand{\ec}{\end{center}}
\newcommand{\bit}{\begin{itemize}}
\newcommand{\eit}{\end{itemize}}
\newcommand{\nn}{\nonumber}

\newcommand{\Sum}{\displaystyle\sum\limits}
\newcommand{\tw}{\textwidth}
\newcommand{\ig}[1]{\includegraphics[width=#1\tw]}

\newcommand{\ssp}{\vspace{0.02\tw}}

\newcommand{\arr}{$\rightarrow$ }

\begin{document}

\title{An algorithm for network community structure determination by surprise}

\date{}

\author[a, *]{Daniel Gamermann}
\author[a]{José Antônio Pellizaro}

\affil[a]{Department of Physics, Universidade Federal do Rio Grande do Sul (UFRGS) - Instituto de Física \\ Av. Bento Gonçalves 9500 - Caixa Postal 15051 - CEP 91501-970 - Porto Alegre, RS, Brasil.}
\affil[*]{Corresponding author e-mail: danielg@if.ufrgs.br}

\maketitle

\begin{abstract}
Graphs representing real world systems may be studied from their underlying community structure. A community in a network is an intuitive idea for which there is no consensus on its objective mathematical definition. The most used metric in order to detect communities is the modularity, though many disadvantages of this parameter have already been noticed in the literature. In this work, we present a new approach based on a different metric: the surprise. Moreover, the biases of different community detection algorithms and benchmark networks are thoroughly studied, identified and commented about.  
\end{abstract}





\section{Introduction}

Many complex systems find straight forward representations through graphs: abstract structures where each vertex corresponds to a given element in the original system and links between these nodes indicate the existence of interactions in the system. Examples for the application of graph theory in the representation of real systems are found in apparently distant fields like molecular biology \cite{bionet2, protnet1}, social sciences \cite{socialnet}, phylogeny \cite{gamermann1}, technology \cite{gridnet1, gridnet2} and many others \cite{flightnet, citenet}.

Graph theory is a well established field in mathematics and properties of these abstract objects do shed light in our understanding of the underlying systems they represent. Given a network, many different parameters, metrics and distributions are defined, many of which measure the same intuitive concepts in different manners \cite{estrada, barabasi}. A particular concept for which there is still no consensus on its precise mathematical definition, though it is a very straightforward intuitive idea, is the concept of community.

A community would be a subset of the network's nodes which are more densely interlinked among them than with the rest of the graph. Though easily stated, there is no clear-cut determined threshold on how densely nodes should be interconnected, or how this threshold might depend on other properties of the vertices or on how big the subset is in order to single out a community in the graph. Many different approaches have been proposed: usually one defines a quality function on the different possible partitions into which a network might be divided and tries to find the one that optimize this function. 

The most widespread quality function used is called modularity \cite{newman2006modularity}, which measures, for a given partition, how much the number of links inside communities deviates from what is expected in a random network with the same number of links. It has been shown, however, that the maximization of this function, the Louvain algorithm \cite{blondel2008fast}, suffers from several pitfalls. The method fails to resolve small communities usually merging them into other communities with which the subset might be only poorly connected (sometimes even unconnected \cite{leiden}). This problem is referred to as a resolution limit \cite{Fortunato36}. On the other hand, modularity presents many similar local maxima, which not only make the finding of the global maximum difficult, but partitions with close modularity values may be discrepant, representing structurally very different partitions of the network into communities \cite{good2010performance}.

Here we present an algorithm based on the maximization of another quality function: the surprise. This function has been presented in \cite{aldecoa2010} in order to assess the quality of partitions found by hierarchical clustering algorithms. Later, the authors proposed a meta-algorithm that, among the different solutions found by different algorithms, selects the one with the highest value of surprise and showed that this meta-algorithm outperforms any of the individual algorithms for known benchmarks used in the field \cite{aldecoa2011deciphering,aldecoa2013surprise}. The implementation of our algorithm and functions used to analyse the results are available as a python package called \texttt{Surpriser} from the python package index website: \texttt{https://pypi.org/project/Surpriser/}.

The work is organized as follows: in the next section the surprise function is presented, along with a description of the algorithm implemented and the benchmarks which are going to be used to compare the performance of our algorithm with others from the literature. In section 3, we present and comment on the obtained results from the application of the different algorithms on the different benchmarks, thoroughly identifying possible biases in the benchmarks and algorithms and then proceed to compare the degeneracy in the modularity and surprise landscapes. Finally, in the last section, we present an overview and our conclusions. In the appendix \ref{appendixPython} we comment on some details of the implementation of the algorithm as a python library and in the appendix \ref{appendixPielou} we explain the Pielou index to measure how homogeneous a division of nodes into communities is.


\section{Algorithms and Benchmarks}

In this section, first we describe the surprise function and explain the workings of our algorithm. Secondly, we present the algorithms against which the results of ours will be compared to and finally all the benchmarks that will be feed to the algorithms.

\subsection{The Surpriser Algorithm}

Given a graph with $K$ nodes and $n$ links, partitioned into $N_c$ communities such that $\ell$ links of the network are between nodes assigned to a same community, the function surprise is defined as\footnote{We chose to work with natural logarithm ($\ln$), though when first presented, the authors originally evaluated the surprise with the base ten logarithm ($\log$).} \cite{aldecoa2010, aldecoa2011deciphering,aldecoa2013surprise}: 

\be
    S &=& - \ln \Sum_{j=\ell}^{\textrm{min($M$,$n$)}} \frac{\binom{M}{j} \binom{F-M}{n-j}}{\binom{F}{n}} .\label{surprise}
\ee
This is a cumulative hyper geometric distribution where, $F$ symbolizes the maximum number of possible links in the network (equivalent to a clique of size $K$: $F=K(K-1)/2$) and $M$ is the maximum possible number of intracommunity links given the partition, such that in a network partitioned into $N_c$ communities where the size of community $i$ is $c_i$, $M$ is given by:

\be
    M &=& \sum_{i=1}^{N_c} \frac{c_i(c_i -1)}{2}.
\ee
One can think of the surprise on measuring how unlikely (surprising) it is to find a partition with as many intracommunity links ($\ell$) as the one in the given graph. Think of an urn with $F$ balls, each ball representing a possible link in the network, where $M$ of these are red, representing possible links inside communities and $F-M$ are blue, representing links between communities. From this urn one extracts $n$ balls, the actual number of links in the graph. The sum in the right-hand side of equation (\ref{surprise}) is the probability of extracting at least $\ell$ red balls \cite{nicolini2016modular} (note that usually $M \ll F$ and therefore the probability of extracting a single red ball might be low).

In previous works \cite{aldecoa2010, aldecoa2011deciphering, aldecoa2013surprise}, the authors used the surprise function in order to select, from the results of different algorithms, the best one, or to tune some algorithm parameter or to choose some cutoff within the algorithms. We choose to tackle this problem head-on developing a greedy algorithm that aims to directly find the partition that maximize the surprise for a network.

The community structure of a network is represented by a partition: to each node in the graph a number between 1 and $N_c$ is assigned indicating to which of the $N_c$ communities the vertex belongs (note that $N_c$ itself may change between different partitions). Two of the parameters on which the surprise function depends on are characteristics of the network as a whole ($F$ and $n$), while the two other parameters are related to the possible partitions of the network alone ($M$ and $\ell$). The surpriser algorithm tests different reassignments of the nodes into communities performing the ones that raise the value of the surprise the most. 

The algorithm starts assigning each node to a different community\footnote{This initial partition could be any but, unless stated otherwise, in all results presented here, in the initial configuration each node was the single element in a community, such that initially the number of nodes and communities are the same.}. A partition may be modified by three basic operations: two communities may be merged into a single one (\texttt{merge}), an element may be exchanged between two communities (\texttt{exchange}) or an element may be removed from an existing community to create a new one\footnote{These last two operations may only be performed over communities with more than one element in it.} (\texttt{extract}). Apart from performing these operations on whole communities and single elements from them, two operations (extract and exchange) may be performed on a ``broader'' scope: given a community, the algorithm runs recursively considering the subgraph composed by the nodes of this community alone, identifying in this way a community's underlying sub-community structure. Then a sub-community may be extracted from its original community or exchanged to another community. These two operations applied on a sub-community level help the algorithm to scape from shallow local maxima.

Supriser is a greedy algorithm, it will follow a loop executing each operation to exhaustion that results in an improved surprise value until no improvement can be found. In the python package implemented, there is also the possibility to execute the operations as an annealing: randomly choosing the communities, the operation performed and accepting the new partition if the surprise is increased or rejecting it with a given probability based on a fictitious temperature if it does not. Our tests, though, show that a more greedy approach resulted in similar (some times even better) results with a much shorter computational time. In a few cases, the only observed gain was applying the annealing operations with a small temperature, after running the greedy algorithm, as to fine tune the optimal solution, but with small difference in the final surprise value. 

The algorithm runs as follows: while any change in the partition has been accepted in a previous loop (or none has been run yet), it runs a new loop over the communities. In each community it runs first a loop over its members and tries first to merge the current community to the one of its members neighbours; if it fails, it tries to exchange an element between the two communities. Once the loop over the community members is finished, it executes the extract operation over the current community until no element can be extracted. Then it runs the subcommunity extraction operation until no improvement is made and finally it makes all possible subcommunity exchanges. The pseudocode for the algorithm can be found in Algorithm \ref{algopc}. In this pseudocode, being successful in an operation means that the operation raised the surprise value for the new partition. The subcommunities of a community are determined applying recursively the algorithm to the subgraph formed by the community elements alone.


\begin{algorithm}
\SetAlgoLined

\KwData{input is a network with a partition;}

 \While{Any change done to the partiton}{
  \For{every community in the partition}{
   \For{every node in the community}{
    \For{every neighbour of the node}{
     \If{community of neighbour $\neq$ community}{
      try to merge communities and if fails,
       try to exchange element between them;}
    }
   }
      \While{successful}{try to extract an element from community;} 
      \While{successful}{try to extract an subcommunity from community;} 
      \While{successful}{
       \For{every other community in partition}{try to relocate a subcommunity to the other community;}} 
     }
 } 
 \caption{Surpriser Algorithm pseudo code. Being successful in an operation means that it increased the surprise value for the updated partition.} \label{algopc}
\end{algorithm}

In regard to the resolution limit, in \cite{aldecoa2013surprise} the authors have already shown that, qualitatively, the surprise does not suffer from a resolution limit in the same way that modularity does. They showed that surprise is able to find the intuitive partition in many networks where modularity maximization fails (for example the ring of cliques network), but no analysis was made about the degeneracy problem. Here we investigate this problem in section \ref{secLandscape} and show that, even though surprise maximization outperforms the other algorithms when dealing with well defined communities (close to cliques), it does suffer from a degeneracy problem (may present many different local, and even global, maxima), but not in the same way as modularity does. Since, the closer two near optimum maxima in the surprise landscape are, the more similar the partitions they represent are, while for modularity, two maxima may be very close in value, but represent structurally very different partitions.

\subsection{Other Algorithms}

We compare the surpriser algorithm results with results from other popular algorithms such as Louvain \cite{blondel2008fast} and the ones implemented in \cite{aldecoa2013}: Constant Potts Model (CPM) \cite{traag2011narrow}, Reichardt and Bornholdt (RB) \cite{reichardt2006statistical}, Ronhovde and Nussinov (RN) \cite{ronhovde2010local}, SCluster \cite{aldecoa2010}, UVCluster \cite{king2004protein} and Infomap \cite{rosvall2008maps}.

The popular Louvain algorithm tries to find a partition that maximizes the modularity of the network. The CPM, RB, and RN are methods based on Potts models which aim to find the ground state of the Hamiltonian of a spin glass associated with the network. Some of these Potts models algorithms have Hamiltonian functions that can be traced back to the modularity by a specific parameter choices; in such cases, the free parameters are a way to try to avoid known short-cuts of the Louvain algorithm, i.e. they parametrize the resolution of the algorithm.

Infomap on the other hand, seeks to minimize the description length of a random walk trough the network links. There are two main ideas behind this: the first is that a random walk over the network links represents the flow of information over that graph and second that communities in this landscape represent groups where the information flows easily and quickly between its members. With that in mind the problem of identifying the community structure boils down to finding an efficient way to describe the paths that arise from the random walks performed in the network. In computational terms, the aim is to find the code that allows one to describe the entirety of a random walk in the least amount of bits, i.e. to minimize the per-step description length of the random walk.

Finally,  UVCluster and SCluster perform an iterative hierarchical clustering in the network. In this context, the concept of hierarchy relates to the existence of communities within communities, that small groups of nodes can be joined to form bigger ones in a structured manner. One way to represent the hierarchy is to construct a dendrogram or a tree of the system. First the actual clustering algorithms are applied on the network then, based on their results, a second algorithm creates the tree (or dendrogram) of the partitions found and cuts this tree at some point in order to identify different branches as different communities.

We should note that, apart from the Louvain algorithm for which we used the implemenation in the python package NetworkX, for the others we use the implementation of these available from the SurpriseMe package \cite{aldecoa2013}. Some of these algorithms have parameters that can be tuned or depend on some user chosen cutoff and, in this implementation, those parameters are set in order to maximize the surprise for the partition returned by the algorithm. Note, though, that the algorithm itself does not maximizes the surprise, but some other quality function that depends on parameters; the implementation then runs the algorithm for different values of these parameters seeking the value that results in the partition with the highest surprise value.


\subsection{Benchmarks}

To test the algorithms, we use three different benchmarks for which the {\it a priori} community structure is known: the Lancichinetti-Fortunato-Radicchi (LFR) \cite{lfrbenchmark}, the relaxed caveman (RC) \cite{aldecoa2010} and a new benchmark (OUR) developed by us that tackles separately different ways in which a community may be degraded: either by loosing internal links or by the creation of links to other groups. This new benchmark will also allow for communities with a single element in it. In this subsection we will first remember the general structure of the LFR and RC benchmarks and then describe our benchmark in more detail.

In the LFR benchmark, the network vertices are generated with a degree distribution that follows a power-law function and then assigned to communities whose sizes distribution is also a scale-free function, though the exponents of the two distributions may be different. The use of scale-free functions for the degree distribution of vertices in a network is inspired by many works in the field \cite{barabasi1, barabasi2, rekaalbert}, though some other works dispute this claim \cite{khanin, gipsi, gamermann2}. The benchmark then connects the nodes among them such that a fraction $\mu$ of each vertex connections will be with nodes from the same community while the remaining fraction $1-\mu$ is with nodes from other communities. Therefore, $\mu$ is the parameter that controls how degraded the a priori community structure is and for high values of $\mu$, one should not expect the benchmarked partition of the graph well representing the actual partition of the network that an algorithm might return. Note that in this benchmark, the connectivity of a node (its degree) is completely independent from its community size; nodes are arbitrarily labeled as belonging to a given community as long as it is possible to connect them to members of this community respecting $\mu$, while the number of nodes with the community label respects the randomly generated community sizes. We interpret such criterion for community assignment as a local one (relative more to the node's characteristics), in contrast to the other two benchmarks that will be discussed next, where communities are first created as cliques and we will refer to such criterion as a global one, for it involves all nodes in the community as a whole.

The Connected Caveman Graphs get their inspiration from the social sciences and are defined by a set of connected cliques. From each clique remove a link and use it to connect a neighboring clique in such a way as to form a ring. The relaxed caveman (RC) benchmark, as proposed in \cite{aldecoa2010}, also starts with a set of cliques, in this case however, they are isolated. Note that this is the strongest possible definition of community: a set of fully connected subgraphs isolated from each other. As identifying these communities should be trivial, the RC benchmark degrades the networks. The degradation process is made in the following way: a percentage $R$ of its links are randomly removed and the same percentage $R$ of the remaining links are shuffled between the nodes of the entire graph. This means that for each network created we have the original structure, that represents the isolated cliques, and a number of additional structures, one for every value of $R$, that represents the network in different stages of the degradation process. The parameter $R$ is, for the RC benchmark, the analogous of what the parameter $\mu$ was in the case of the LFR benchmark. For the generation of the initial clique sizes in this benchmark, one sets a value for the distribution's Pielou index parameter \cite{pielou}. This parameter is a measure of how even (similar) numbers in a set are. In this case, the numbers are the communities sizes and a pielou index close to 1 (its maximum possible value) would indicate a homogeneous community structure (all communities with similar sizes) while a small value (close to zero) indicates very heterogeneous community structure. In appendix \ref{appendixPielou} we briefly explain in more detail the evaluation and interpretation of this index.

Finally, we propose a new benchmark also inspired by the idea that a graph perfectly divided into communities would be a set of disconnected cliques. The aforementioned benchmarks define a single parameter as a measure of how degraded the community structure of the network is ($\mu$ or $R$). We propose two parameters in order to control the different ways in which this perfect community structure might be disrupted: a connection inside a clique might be lost with probability $p$ and each possible connection between cliques may appear with probability $q$. Moreover, cliques (communities) in the LFR and RC benchmarks must have at least 2 elements (graph nodes) in it, we propose to build a network where a fraction $r$ of its nodes is assigned to its own community alone. 

The creation of OUR benchmark network follows the steps:

\bi
\item Creation of a given number of cliques. In the present work, the cliques sizes will be set in order to achieve a given pielou index, as in the RC benchmark.
\item Each link inside each clique $i$ with size $c_i$ may be lost with probability $p$.
\item Given two cliques of sizes $c_i$ and $c_j$, each one of the $c_i c_j$ possible connections between the two is created with probability $q$.
\item A fraction $r$ of nodes is randomly connected to the network (each one of these is alone in its own community, with a single connection to a randomly chosen clique).
\ei

So the benchmark network starts from a sequence of $n_c$ clique sizes: $c_1, c_2, ..., c_{n_c}$. The resulting total number of nodes in such a network is:

\be
K &=& \left\lfloor\frac{1}{(1-r)}\Sum_{i=1}^{n_c} c_i\right\rfloor,
\ee
the total number of communities in the network is $N_c=n_c+\lfloor rK\rfloor$, the expected number of links inside communities is

\be
\langle n_{in} \rangle &=& (1-p)\Sum_{i=1}^{n_c}\frac{c_i(c_i-1)}{2}
\ee
and the expected number of links between communities is 

\be
\langle n_{out} \rangle &=& q\Sum_{i=1}^{N_c}\Sum_{j<i}^{N_c}c_i c_j + \frac{r}{1-r}\Sum_{i=1}^{n_c}c_i.
\ee

We shall comment here that in the implementation of this benchmark in the python package, the probabilities $p$ and $q$ may be functions of the cliques sizes, though in the present work we used constant values in order to keep handling a small number of parameters. But it is important to realize that bigger values of $p$ will have much more impact in smaller cliques and therefore, unless one smartly scales $p$ with the cliques sizes, the communities will not be uniformly degraded (smaller communities are degraded faster). This is an issue in the other benchmarks as well. In fact, in the RC graphs, for example, links are shuffled and removed, such that for high values of $R$ the connection density of the graph might be very low and some nodes can even get disconnected from the network.

Finally we note that in order to compare the algorithms results with the initial community partitions given by the benchmarks, we use the Variation of Information (VI) parameter \cite{vi}. This concept, from information theory (similarly to the Pielou index), measures the similarity between two lists as the normalized joint entropy between the information contained in the lists. The lists here are the partitions: $K$ numbers between 1 and the number of communities in the partition indicating to which community each node in a network belongs. A VI equal to 0 indicates that the two compared partitions have the same information (represent the same community structure) while a value of 1 indicates completely different community divisions.


\section{Results}

In this section we describe first all the benchmark graphs that were generated. Next, we present the results of running the different algorithms in each benchmark and analyze their differences and possible biases. In the last subsection, we tackle the degeneracy problem by visualising the surprise and modularity landscapes for a very simple graph. 

\subsection{Generated Benchmarks}

For each benchmark, we generated two sets of networks, one where the graphs have 500 nodes each, and another with 1000 nodes each. Each set, for each benchmark, generated with different parameters, is comprised of 1000 graphs, where 100 of each is with a different degradation level. We refer to these two sets as Big and Small (note that the sets are comprised from the same number of elements, what is bigger or smaller are the graphs within each set). 

In order to present the results of the three benchmarks in an uniform fashion, we use the symbol $\mu$, which can have values between 0 and 1, when referring to the degradation level of the community structure in all benchmarks. In LFR, it stands for the $\mu$ already explained in the previous section, while in the RC benchmark, since $R$ was referred to as a percentage, it is straight forward to set $\mu=R/100$. In OUR benchmark, there are two parameters that control the degradation ($p$ and $q$), and we set $p=\mu$ and $q=0.05\mu$, such that to each value of $\mu$ there are corresponding values of $p$ and $q$. The value of $q$ is chosen to be smaller than $p$ because the number of possible links between communities is more numerous than the possible links inside communities.

For the other parameters in each benchmark, we made the following choices explained below.

In the RC benchmark, one must set the number of communities and the pielou index for the set of clique sizes. For the Small set we chose $N_c=20$, while in the Big set $N_c=40$. The clique list was generated with three possible values for the pielou index, $PI=0.75, 0.85$ or $0.95$. Note however, that sometimes it is hard or even not possible to generate numbers that lead to the exact desired value for the pielou index and therefore, the distribution of the pielou values generated has a spread around its goal value.

The average degree of the nodes for the LFR graphs is set to $\bar{k}=15$, which is a value usually used in the literature \cite{lfrbenchmark, orman2009comparison}. For the exponent in the power-law distribution for the nodes degree ($p(k)\propto k^{-\gamma}$), the value $\gamma=2$ is usually chosen, but we shall note here that this value might be incompatible with the choice made for $\bar{k}$. Given the power-law distribution, one might evaluate its average (expected value in the limit $K \to \infty$) as:

\be
\langle k \rangle &=& \frac{\zeta(\gamma-1, 1)}{\zeta(\gamma, 1)}, \label{kexpected} \\
\zeta(\gamma, x_0) &=& \Sum_{k=x_0}^{\infty} k^{-\gamma} \label{riemannzeta}
\ee
where $\zeta(\gamma, x_0)$ is the Riemann zeta function (modified to allow the sum to star at any natural number, $x_0$) and $\gamma$ is the power-law exponent. Moreover, another parameter that must be feed into the benchmark is the maximum possible degree of a node and, if badly chosen, the probability of generating a number bigger than $k_{max}$ might be such that, when generating $K$ numbers (degrees for all vertices), a sizable fraction of them might have $k>k_{max}$ according to:

\be
p(k> k_{max}) &=&  \frac{\zeta(\gamma, k_{max}+1)}{\zeta(\gamma, 1)}.
\ee
Indeed, as mentioned, typical values used in the literature \cite{lfrbenchmark, orman2009comparison} are $\gamma=2$, $\bar{k}=15$ and $k_{max}=45$ for $K=1000$. For $\gamma=2$, equation (\ref{kexpected}) diverges but the value of $\gamma$ compatible with $\langle k \rangle=15$ is close ($\gamma=2.0425$) for which $K p(k>45)=11.15$ meaning that when generating 1000 numbers, around 11 (13 if one uses $\gamma=2$) will have to be discarded for having a degree bigger than the maximum allowed. In \cite{traag2011narrow, aldecoa2012closed} the authors use $\bar{k}=20$, for which the corresponding $\gamma$ would be $2.0315$, and $k_{max}=50$ which would result in around 10 degrees having to be discarded. In the graphs generated for this work, we kept the value $\bar{k}=15$, but use $\gamma=2.0425$ and $k_{max}=250$, instead. This benchmark also asks for the exponent for the power-law for setting the community sizes that was set to $\beta=1$ and the maximum community size was chosen to be 300. Note that the pielou index for the communities sizes in this benchmark is only indirectly controlled via the $\beta$ parameter (exponent for the community sizes distribution).

Finally, in OUR benchmark we set the parameter $r=0.01$ (1\% of the nodes will be alone in their own communities). In the Small set, we generated 20 cliques, while for $K=1000$ we generated 40 cliques\footnote{Note that since 1\% of the network nodes are each alone in their own communities and the cliques contain the other 99\% of nodes, the number of communities in the Small and Big sets are respectively 25 and 50.}. In each set of cliques, the numbers chosen for the clique sizes were such that the pielou index for the communities sizes in each graph was either $PI=0.75, 0.85$ or $0.95$, but as in the RC case, one must expect a spread in the values distribution around these goals.

In total, considering the different sets of parameters, in the different benchmarks a total of 14000 graphs were produced. To illustrate all the benchmark networks that are feed to the algorithms, Table \ref{tabbenchs} summarizes the number of graphs produced for each benchmark with the different parameters and characteristics.

\btab[h]
\caption{Number of benchmark graphs generated for the different values of the parameters. The column Pielou show the average and standard deviation for the pielou indexes for the list of community sizes generated by the benchmark. The values $PI=0.75, 0.85, 0.95$ mentioned in the text for the RC and OUR benchmarks are not exactly reached by the random generator algorithms and in the LFR benchmark, the pielou index for the community sizes can be only indirectly controlled via the $\beta$ parameter.}\label{tabbenchs}
\bc
\footnotesize
\bt{c||c|c||cccccccccc|c}
\multirow{2}{*}{Benchmark}  &  \multirow{2}{*}{$K$}   &  \multirow{2}{*}{Pielou}    & \multicolumn{10}{c|}{$\mu$} & \multirow{2}{*}{Total} \\
\cline{4-13}
  &  &  & 0.0 &  0.1 &  0.2 &  0.3 &  0.4 &  0.5 &  0.6 &  0.7 &  0.8 &  0.9 &  \\
\hline
\hline
\multirow{2}{*}{LFR} & 500 & 0.9282 $\pm$ 0.0747 & 100 & 100 & 100 & 100 & 100 & 100 & 100 & 100 & 100 & 100 & 1000 \\
\cline{2-14}
 & 1000 & 0.9152 $\pm$ 0.0802 & 100 & 100 & 100 & 100 & 100 & 100 & 100 & 100 & 100 & 100 & 1000 \\
\hline
\multirow{6}{*}{RC} & \multirow{3}{*}{500} & 0.7505 $\pm$ 0.0030 & 100 & 100 & 100 & 100 & 100 & 100 & 100 & 100 & 100 & 100 & 1000 \\
\cline{3-14}
 &  & 0.8497 $\pm$ 0.0029 & 100 & 100 & 100 & 100 & 100 & 100 & 100 & 100 & 100 & 100 & 1000 \\
\cline{3-14}
 &  & 0.9491 $\pm$ 0.0028 & 100 & 100 & 100 & 100 & 100 & 100 & 100 & 100 & 100 & 100 & 1000 \\
\cline{3-14}
\cline{2-14}
 & \multirow{3}{*}{1000} & 0.7768 $\pm$ 0.0135 & 100 & 100 & 100 & 100 & 100 & 100 & 100 & 100 & 100 & 100 & 1000 \\
\cline{3-14}
 &  & 0.8507 $\pm$ 0.0028 & 100 & 100 & 100 & 100 & 100 & 100 & 100 & 100 & 100 & 100 & 1000 \\
\cline{3-14}
 &  & 0.9482 $\pm$ 0.0026 & 100 & 100 & 100 & 100 & 100 & 100 & 100 & 100 & 100 & 100 & 1000 \\
\cline{3-14}
\hline
\multirow{6}{*}{OUR} & \multirow{3}{*}{500} & 0.7531 $\pm$ 0.0005 & 100 & 100 & 100 & 100 & 100 & 100 & 100 & 100 & 100 & 100 & 1000 \\
\cline{3-14}
 &  & 0.8531 $\pm$ 0.0005 & 100 & 100 & 100 & 100 & 100 & 100 & 100 & 100 & 100 & 100 & 1000 \\
\cline{3-14}
 &  & 0.9433 $\pm$ 0.0002 & 100 & 100 & 100 & 100 & 100 & 100 & 100 & 100 & 100 & 100 & 1000 \\
\cline{3-14}
\cline{2-14}
 & \multirow{3}{*}{1000} & 0.7525 $\pm$ 0.0005 & 100 & 100 & 100 & 100 & 100 & 100 & 100 & 100 & 100 & 100 & 1000 \\
\cline{3-14}
 &  & 0.8526 $\pm$ 0.0005 & 100 & 100 & 100 & 100 & 100 & 100 & 100 & 100 & 100 & 100 & 1000 \\
\cline{3-14}
 &  & 0.9525 $\pm$ 0.0005 & 100 & 100 & 100 & 100 & 100 & 100 & 100 & 100 & 100 & 100 & 1000 \\
\cline{3-14}
\hline
\hline
Total & \multicolumn{2}{c|}{ ~ } & 1400 & 1400 & 1400 & 1400 & 1400 & 1400 & 1400 & 1400 & 1400 & 1400 & 14000
\et
\ec
\etab

\subsection{Algorithms Performance}

Each one of the 8 algorithms (Louvain, CPM, Infomap, RB, RN, SCluster, UVCluster and Surpriser) was executed over each one of the 14000 benchmarked networks. The partition returned by each algorithm was compared with the initial partition (the benchmarked one used to generate the graph in the first place) by evaluating the variation of information (VI) between the two. In figure \ref{fig1} we present a global result showing the histograms for the average VI obtained over all networks together.

\bfig
\bc
\ig{0.7}{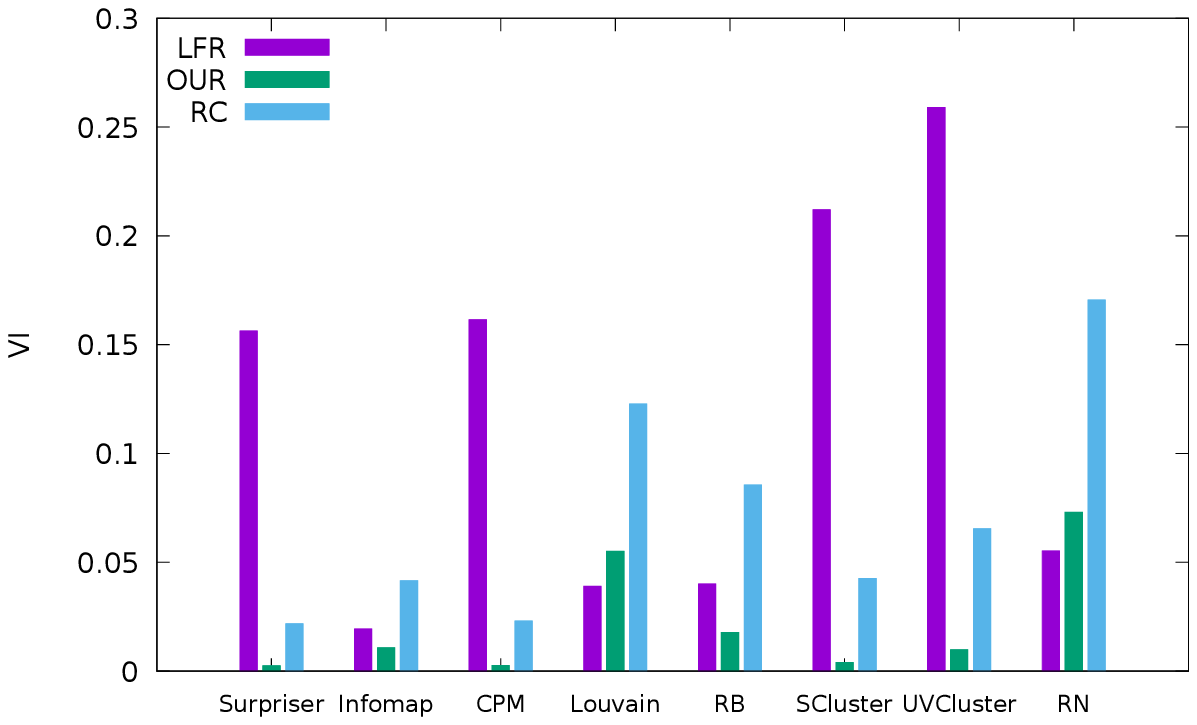}

\ig{0.7}{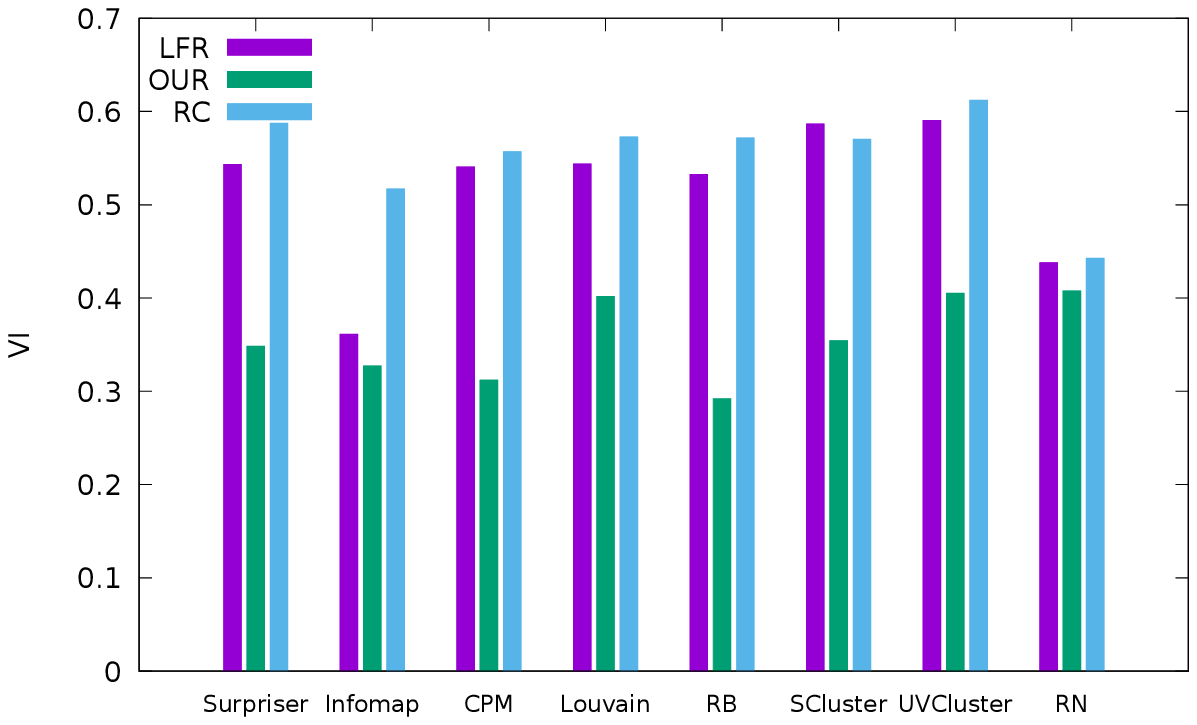}

\ig{0.7}{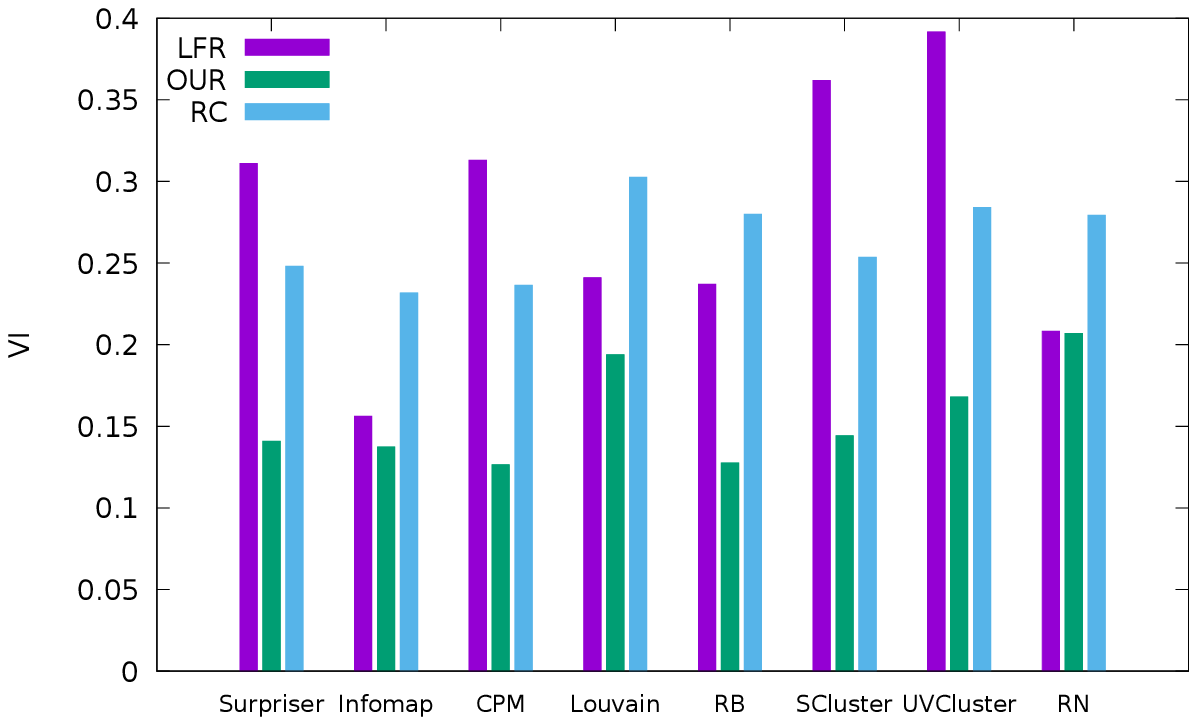}
\ec
\caption{Average variation of information between the benchmark partitions and the ones returned by the algorithms. Upper plot: only those for which $\mu\le0.5$. Middle plot: only those for which $\mu>0.5$. Bottom plot: all values of $\mu$ together.}\label{fig1}
\efig

In figure \ref{fig2} we show the same bar plots, but separating those benchmarks from the Small and Big sets (with $K=500$ and $K=1000$, respectively). From this figure one sees that only the Louvain and RN algorithms seem to behave differently in some benchmarks given the different sizes of the graphs. Finally, in Table \ref{tabresults} we show the numerical VI statics for each algorithm in each set and in both sets together.

\bfig
\bc
\bt{cc}
\ig{0.45}{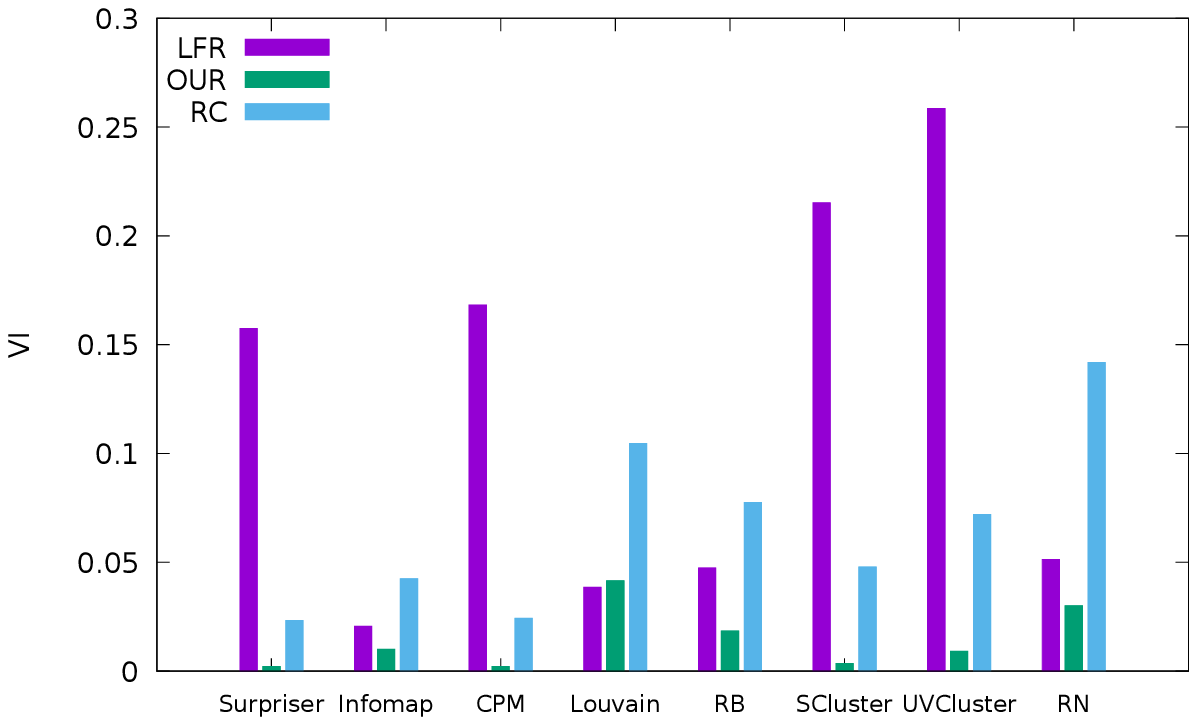} & \ig{0.45}{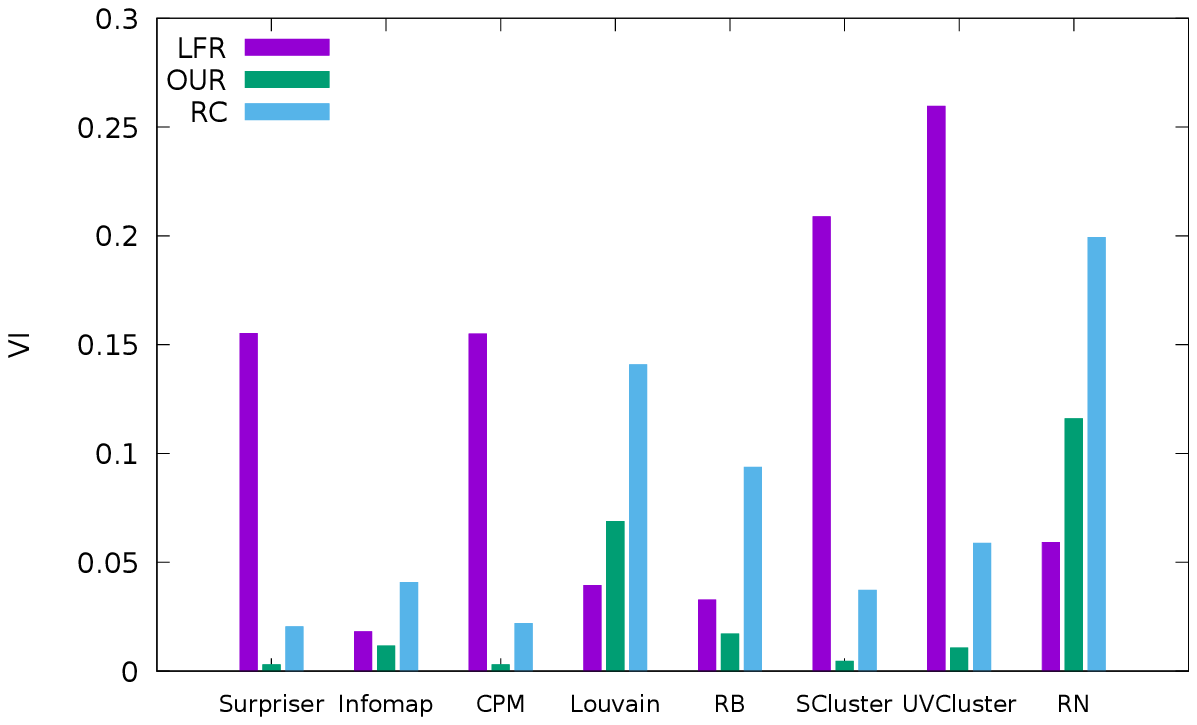} \\
\ig{0.45}{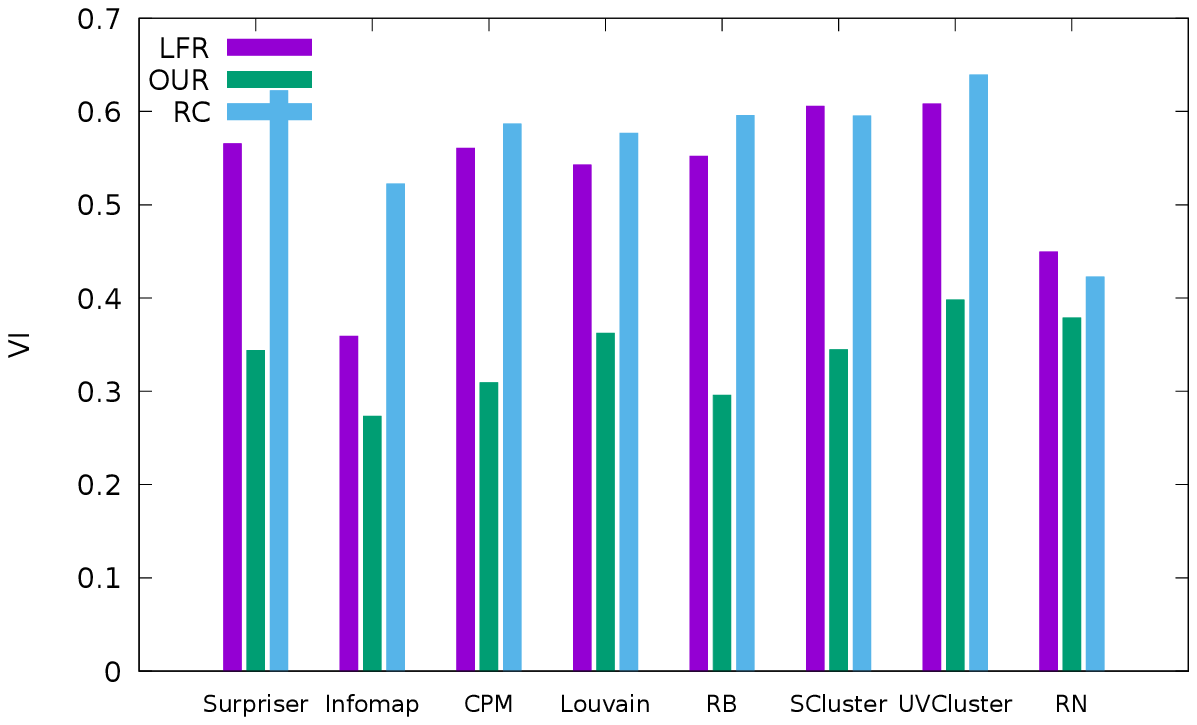} & \ig{0.45}{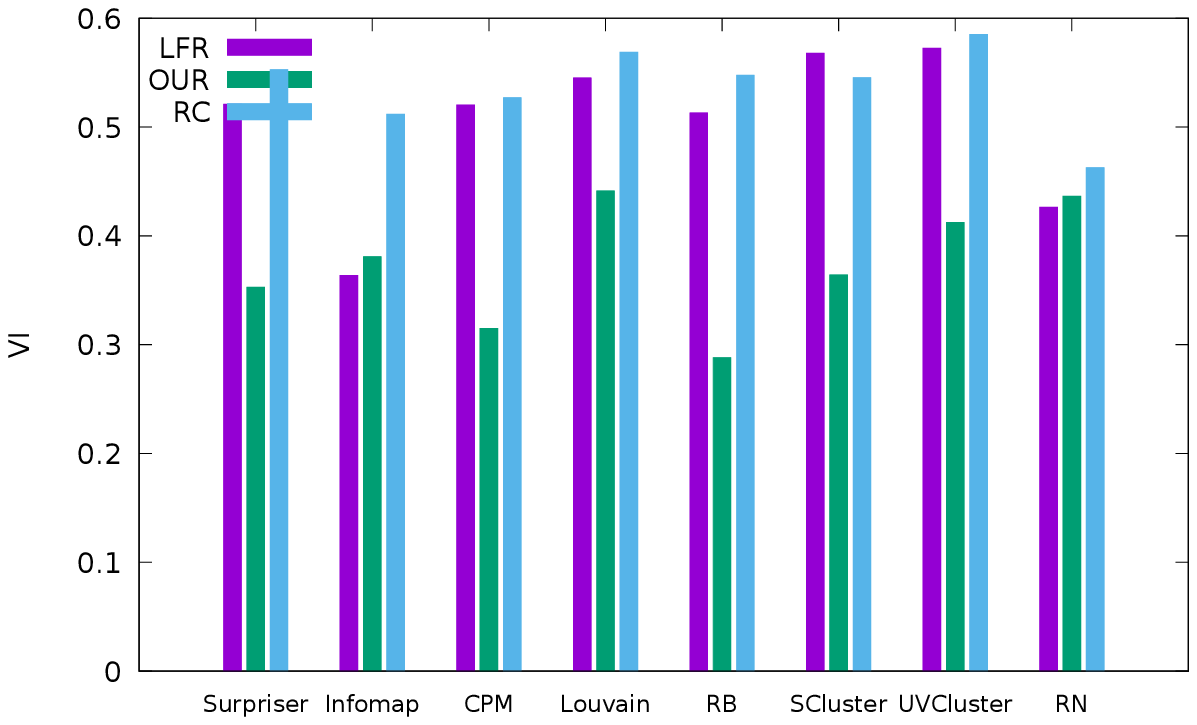} \\
\ig{0.45}{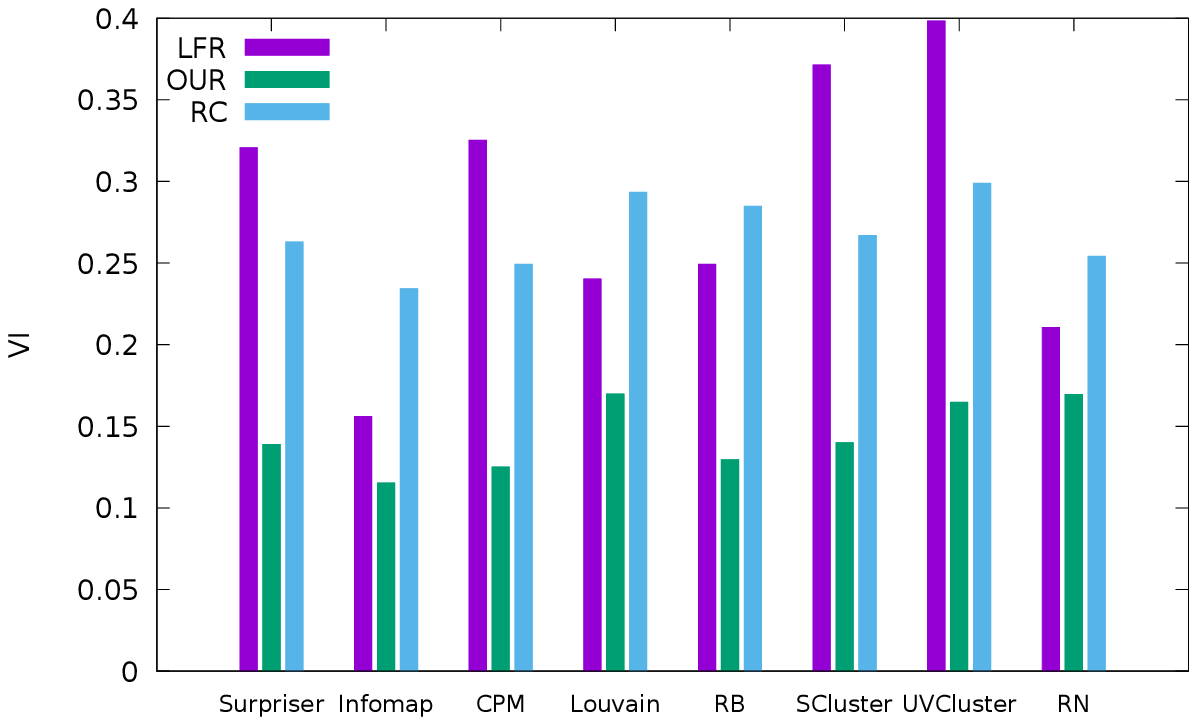} & \ig{0.45}{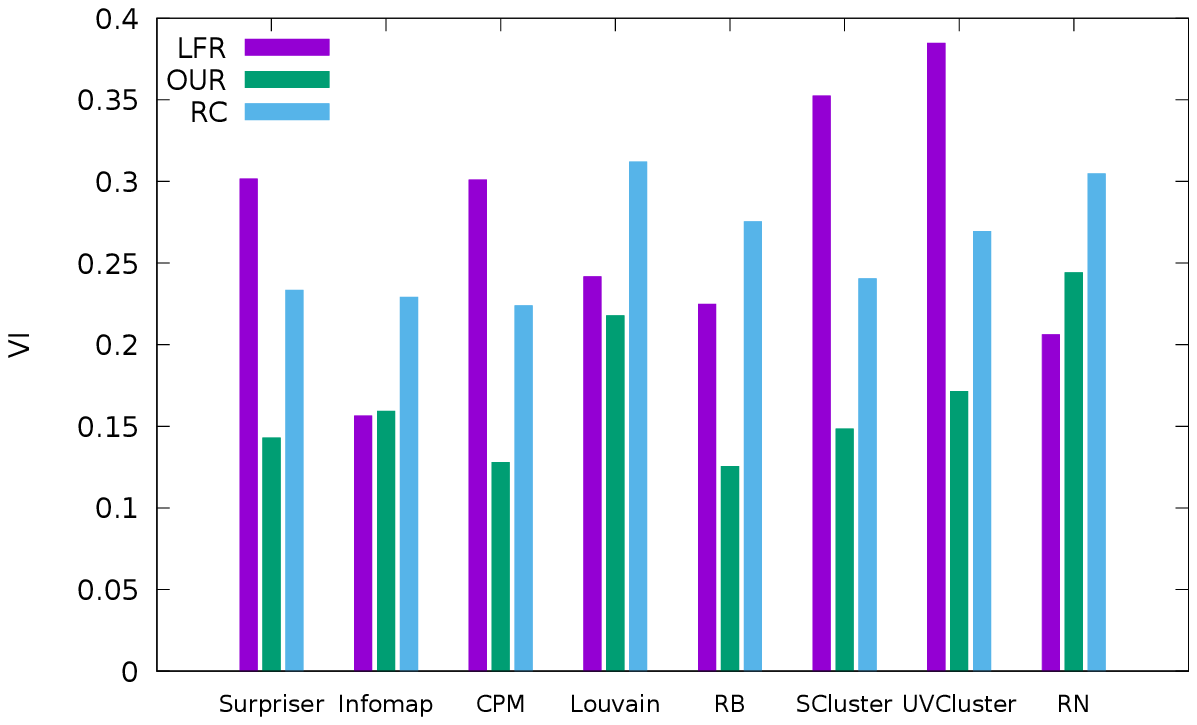}
\et
\ec
\caption{Average variation of information between the benchmark partitions and the ones returned by the algorithms. In the left results from the Small set and on the right for the Big set. Upper plots: only those for which $\mu\le0.5$. Middle plots: only those for which $\mu>0.5$. Bottom plots: all values of $\mu$ together.}\label{fig2}
\efig

\bgroup
\def\arraystretch{1.5}
\begin{sidewaystable}
\caption{Average variation of information between each algorithm returned partition and the benchmarked one for each set and both sets together.}\label{tabresults}
\bc
\scriptsize
\bt{c|c|ccc|ccc|ccc}
 & \multirow{2}{*}{Algorithm} & \multicolumn{3}{c}{$\mu\le0.5$} & \multicolumn{3}{|c}{$\mu>0.5$} & \multicolumn{3}{|c}{$\forall\mu$} \\
 \cline{3-11}
 & & LFR & OUR & RC & LFR & OUR & RC & LFR & OUR & RC \\
\hline
\hline
\multirow{8}{*}{\rotatebox{90}{ K = all }}
 & surpriser  & $0.156224^{+0.347973}_{-0.138985}$ & $0.002512^{+0.005557}_{-0.002129}$ & $0.021792^{+0.065163}_{-0.019022}$ & $0.543164^{+0.142236}_{-0.285633}$ & $0.348258^{+0.312608}_{-0.291510}$ & $0.587452^{+0.113641}_{-0.265537}$ & $0.311000^{+0.308698}_{-0.265995}$ & $0.140810^{+0.475892}_{-0.132232}$ & $0.248056^{+0.403397}_{-0.222464}$ \\
\cline{2-11}
 & Infomap    & $0.019352^{+0.131604}_{-0.018865}$ & $0.010768^{+0.010649}_{-0.004393}$ & $0.041587^{+0.158684}_{-0.035142}$ & $0.361164^{+0.141570}_{-0.252211}$ & $0.326995^{+0.149816}_{-0.277947}$ & $0.516828^{+0.207121}_{-0.191159}$ & $0.156077^{+0.293841}_{-0.147625}$ & $0.137259^{+0.333862}_{-0.121256}$ & $0.231684^{+0.355930}_{-0.203272}$ \\
\cline{2-11}
 & CPM        & $0.161471^{+0.403684}_{-0.143595}$ & $0.002531^{+0.006040}_{-0.002218}$ & $0.023060^{+0.066358}_{-0.020158}$ & $0.540275^{+0.135456}_{-0.273907}$ & $0.311993^{+0.342977}_{-0.248997}$ & $0.556604^{+0.136446}_{-0.268127}$ & $0.312993^{+0.306875}_{-0.275028}$ & $0.126316^{+0.461834}_{-0.116990}$ & $0.236477^{+0.395297}_{-0.209680}$ \\
\cline{2-11}
 & louvain    & $0.038935^{+0.118429}_{-0.029560}$ & $0.055122^{+0.044741}_{-0.037231}$ & $0.122681^{+0.100141}_{-0.087719}$ & $0.543699^{+0.176514}_{-0.275586}$ & $0.401635^{+0.286015}_{-0.231044}$ & $0.572529^{+0.137736}_{-0.222050}$ & $0.240841^{+0.397118}_{-0.208311}$ & $0.193727^{+0.399163}_{-0.134548}$ & $0.302620^{+0.327799}_{-0.200599}$ \\
\cline{2-11}
 & RB         & $0.039999^{+0.260641}_{-0.036716}$ & $0.017690^{+0.020878}_{-0.012574}$ & $0.085540^{+0.125923}_{-0.067498}$ & $0.532338^{+0.168059}_{-0.302635}$ & $0.291891^{+0.332545}_{-0.230356}$ & $0.571577^{+0.109573}_{-0.270465}$ & $0.236935^{+0.390923}_{-0.217762}$ & $0.127370^{+0.451796}_{-0.104400}$ & $0.279955^{+0.351640}_{-0.211402}$ \\
\cline{2-11}
 & SCluster   & $0.211948^{+0.328906}_{-0.158693}$ & $0.004021^{+0.008730}_{-0.003296}$ & $0.042482^{+0.090867}_{-0.033882}$ & $0.586641^{+0.113106}_{-0.169615}$ & $0.354275^{+0.287687}_{-0.245940}$ & $0.570229^{+0.114065}_{-0.200998}$ & $0.361825^{+0.269044}_{-0.272823}$ & $0.144123^{+0.415861}_{-0.131795}$ & $0.253581^{+0.352204}_{-0.218460}$ \\
\cline{2-11}
 & UVCluster  & $0.258975^{+0.284361}_{-0.166975}$ & $0.009847^{+0.017529}_{-0.007602}$ & $0.065389^{+0.133929}_{-0.051315}$ & $0.590155^{+0.105846}_{-0.129907}$ & $0.405017^{+0.224273}_{-0.261369}$ & $0.612014^{+0.076594}_{-0.139285}$ & $0.391447^{+0.238701}_{-0.254478}$ & $0.167915^{+0.398952}_{-0.148320}$ & $0.284039^{+0.336954}_{-0.238197}$ \\
\cline{2-11}
 & RN         & $0.055192^{+0.268569}_{-0.054517}$ & $0.072948^{+0.365014}_{-0.071154}$ & $0.170522^{+0.259323}_{-0.168346}$ & $0.437742^{+0.114728}_{-0.294709}$ & $0.407422^{+0.082133}_{-0.257241}$ & $0.442545^{+0.064711}_{-0.052193}$ & $0.208212^{+0.289893}_{-0.198191}$ & $0.206738^{+0.253272}_{-0.203372}$ & $0.279331^{+0.166243}_{-0.277138}$ \\
\cline{2-11}
\hline
\multirow{8}{*}{\rotatebox{90}{ K = 500 }}
 & surpriser  & $0.157415^{+0.419289}_{-0.135585}$ & $0.002062^{+0.005377}_{-0.001771}$ & $0.023185^{+0.068211}_{-0.020304}$ & $0.565423^{+0.140053}_{-0.262129}$ & $0.343760^{+0.318323}_{-0.288472}$ & $0.622210^{+0.092908}_{-0.219458}$ & $0.320618^{+0.320594}_{-0.277342}$ & $0.138741^{+0.478502}_{-0.130513}$ & $0.262795^{+0.394890}_{-0.240420}$ \\
\cline{2-11}
 & Infomap    & $0.020637^{+0.143979}_{-0.020084}$ & $0.010024^{+0.009387}_{-0.003729}$ & $0.042399^{+0.153844}_{-0.036120}$ & $0.358807^{+0.124365}_{-0.256528}$ & $0.273287^{+0.177103}_{-0.222920}$ & $0.522031^{+0.201107}_{-0.201629}$ & $0.155905^{+0.281451}_{-0.148612}$ & $0.115329^{+0.326715}_{-0.099299}$ & $0.234252^{+0.355106}_{-0.206544}$ \\
\cline{2-11}
 & CPM        & $0.168121^{+0.441743}_{-0.144282}$ & $0.002100^{+0.006214}_{-0.001863}$ & $0.024284^{+0.069536}_{-0.021280}$ & $0.560523^{+0.132445}_{-0.252063}$ & $0.309142^{+0.348528}_{-0.249137}$ & $0.586443^{+0.114131}_{-0.312212}$ & $0.325082^{+0.316900}_{-0.279261}$ & $0.124917^{+0.470483}_{-0.115712}$ & $0.249148^{+0.401938}_{-0.222659}$ \\
\cline{2-11}
 & louvain    & $0.038562^{+0.177979}_{-0.030219}$ & $0.041484^{+0.036497}_{-0.028441}$ & $0.104574^{+0.084178}_{-0.075721}$ & $0.542460^{+0.167581}_{-0.283240}$ & $0.362124^{+0.312447}_{-0.215752}$ & $0.576557^{+0.123927}_{-0.237577}$ & $0.240121^{+0.393210}_{-0.213166}$ & $0.169740^{+0.400995}_{-0.122329}$ & $0.293367^{+0.349046}_{-0.198836}$ \\
\cline{2-11}
 & RB         & $0.047356^{+0.327808}_{-0.043706}$ & $0.018428^{+0.024810}_{-0.012759}$ & $0.077454^{+0.112927}_{-0.061861}$ & $0.551837^{+0.170097}_{-0.287961}$ & $0.295821^{+0.329239}_{-0.234265}$ & $0.595586^{+0.097220}_{-0.264811}$ & $0.249148^{+0.397968}_{-0.228545}$ & $0.129385^{+0.451367}_{-0.106193}$ & $0.284707^{+0.368580}_{-0.217572}$ \\
\cline{2-11}
 & SCluster   & $0.215129^{+0.330366}_{-0.162063}$ & $0.003472^{+0.008400}_{-0.002918}$ & $0.047894^{+0.095435}_{-0.038540}$ & $0.605544^{+0.109036}_{-0.173235}$ & $0.344590^{+0.301466}_{-0.238474}$ & $0.595254^{+0.100620}_{-0.210142}$ & $0.371295^{+0.276666}_{-0.279852}$ & $0.139919^{+0.415732}_{-0.128414}$ & $0.266838^{+0.359431}_{-0.227631}$ \\
\cline{2-11}
 & UVCluster  & $0.258443^{+0.278190}_{-0.166150}$ & $0.009080^{+0.016451}_{-0.007090}$ & $0.071972^{+0.138750}_{-0.056203}$ & $0.607876^{+0.097667}_{-0.129180}$ & $0.397871^{+0.241048}_{-0.254188}$ & $0.639056^{+0.066783}_{-0.108886}$ & $0.398216^{+0.241564}_{-0.259479}$ & $0.164597^{+0.399917}_{-0.146384}$ & $0.298806^{+0.343666}_{-0.247831}$ \\
\cline{2-11}
 & RN         & $0.051251^{+0.267934}_{-0.050448}$ & $0.030010^{+0.383810}_{-0.028300}$ & $0.141789^{+0.257954}_{-0.139166}$ & $0.449263^{+0.125568}_{-0.291267}$ & $0.378554^{+0.076795}_{-0.345355}$ & $0.422532^{+0.089333}_{-0.044170}$ & $0.210456^{+0.312596}_{-0.194690}$ & $0.169428^{+0.270507}_{-0.166306}$ & $0.254086^{+0.169737}_{-0.251434}$ \\
\cline{2-11}
\hline
\multirow{8}{*}{\rotatebox{90}{ K = 1000 }}
 & surpriser  & $0.155033^{+0.290135}_{-0.143134}$ & $0.002961^{+0.005623}_{-0.002463}$ & $0.020399^{+0.061620}_{-0.017770}$ & $0.520905^{+0.144553}_{-0.302535}$ & $0.352757^{+0.308057}_{-0.293958}$ & $0.552694^{+0.130756}_{-0.297033}$ & $0.301382^{+0.292826}_{-0.257102}$ & $0.142879^{+0.471927}_{-0.134125}$ & $0.233317^{+0.410378}_{-0.205359}$ \\
\cline{2-11}
 & Infomap    & $0.018067^{+0.118485}_{-0.017716}$ & $0.011512^{+0.011565}_{-0.005048}$ & $0.040776^{+0.164233}_{-0.034173}$ & $0.363521^{+0.159289}_{-0.247119}$ & $0.380704^{+0.114990}_{-0.335227}$ & $0.511626^{+0.212816}_{-0.182772}$ & $0.156248^{+0.306378}_{-0.146887}$ & $0.159189^{+0.329628}_{-0.144040}$ & $0.229116^{+0.356925}_{-0.200091}$ \\
\cline{2-11}
 & CPM        & $0.154821^{+0.372743}_{-0.142472}$ & $0.002962^{+0.005889}_{-0.002539}$ & $0.021835^{+0.063066}_{-0.019028}$ & $0.520027^{+0.138427}_{-0.290286}$ & $0.314845^{+0.338130}_{-0.248764}$ & $0.526765^{+0.153977}_{-0.240518}$ & $0.300903^{+0.294646}_{-0.272321}$ & $0.127715^{+0.453172}_{-0.118344}$ & $0.223807^{+0.385322}_{-0.197232}$ \\
\cline{2-11}
 & louvain    & $0.039308^{+0.071047}_{-0.028496}$ & $0.068759^{+0.043946}_{-0.044080}$ & $0.140787^{+0.105146}_{-0.100678}$ & $0.544938^{+0.185767}_{-0.269192}$ & $0.441145^{+0.253162}_{-0.243397}$ & $0.568501^{+0.152523}_{-0.207143}$ & $0.241560^{+0.401699}_{-0.203577}$ & $0.217714^{+0.395289}_{-0.144400}$ & $0.311873^{+0.312719}_{-0.199432}$ \\
\cline{2-11}
 & RB         & $0.032643^{+0.188837}_{-0.029645}$ & $0.016951^{+0.016730}_{-0.012470}$ & $0.093625^{+0.136150}_{-0.073238}$ & $0.512840^{+0.166355}_{-0.313352}$ & $0.287961^{+0.335420}_{-0.226917}$ & $0.547567^{+0.122521}_{-0.257831}$ & $0.224722^{+0.381793}_{-0.207330}$ & $0.125355^{+0.449863}_{-0.102780}$ & $0.275202^{+0.333534}_{-0.205560}$ \\
\cline{2-11}
 & SCluster   & $0.208766^{+0.323935}_{-0.156551}$ & $0.004570^{+0.008822}_{-0.003671}$ & $0.037069^{+0.084059}_{-0.029053}$ & $0.567739^{+0.106693}_{-0.182435}$ & $0.363961^{+0.281870}_{-0.247879}$ & $0.545204^{+0.123722}_{-0.191209}$ & $0.352355^{+0.261432}_{-0.265228}$ & $0.148326^{+0.414211}_{-0.135405}$ & $0.240323^{+0.342210}_{-0.209667}$ \\
\cline{2-11}
 & UVCluster  & $0.259507^{+0.290671}_{-0.168238}$ & $0.010615^{+0.018483}_{-0.008088}$ & $0.058806^{+0.126945}_{-0.046284}$ & $0.572433^{+0.108307}_{-0.133139}$ & $0.412162^{+0.208231}_{-0.270252}$ & $0.584972^{+0.083961}_{-0.152343}$ & $0.384678^{+0.234667}_{-0.250762}$ & $0.171234^{+0.399088}_{-0.150092}$ & $0.269272^{+0.328686}_{-0.228269}$ \\
\cline{2-11}
 & RN         & $0.059134^{+0.269827}_{-0.058645}$ & $0.115886^{+0.328279}_{-0.114160}$ & $0.199254^{+0.252501}_{-0.197704}$ & $0.426221^{+0.100752}_{-0.299561}$ & $0.436290^{+0.084421}_{-0.197672}$ & $0.462557^{+0.059193}_{-0.034840}$ & $0.205969^{+0.269078}_{-0.201770}$ & $0.244048^{+0.230341}_{-0.240416}$ & $0.304576^{+0.158272}_{-0.303059}$ \\
\cline{2-11}
\hline
\et
\ec
\end{sidewaystable}
\egroup

One should expect that as $\mu$ increases, the more unlikely it is for the algorithms to resolve the initial benchmark communities, since the initial community assignment made by the benchmark will be more and more degraded. Moreover, one can expect that smaller communities will be broken faster than bigger ones (a point we will comment further when analysing the bias the pielou index has in the algorithms). So, in order to comment on the algorithms performance, let's focus on the top plot of figure \ref{fig1}, where only benchmarks for which $\mu\le0.5$ have been considered.

The first thing to notice is a dichotomy: algorithms that seem to perform better for the LFR benchmark, usually perform worst for the RC and OUR benchmarks and vice-versa. This certainly points to different biases in the algorithms, but it might also be interpreted as a glitch in the community assignment made by the benchmarks in the first place.

In order to understand the discrepancies between each algorithm returned partition and the ones generated by the benchmarks, let's analyse the overlap between the communities in the initial partition produced by the benchmarks and the community assignments made by the algorithms. Nodes in a given initial community may be distributed among different communities by a given algorithm. We will differentiate three cases here: i) in a community identified by the algorithm, a number of nodes coming from the same initial community represents more than 50\% of this initial community size. In this case, we say the community identity as a whole was preserved by such fragment. ii) in a community identified by the algorithm, a fraction of nodes from a same initial community is bigger or equal to the fraction of nodes from different initial communities in the algorithm identified group, but less than 50\% of the initial community size. This is then a fragment, though not one that preserves the initial community identity. iii) in any other case, the nodes from a given initial community in the algorithm determined groups have lost their identities and were dispersed. Considering these three cases, it is possible to classify each network node in an algorithm's partition as a node that partially {\bf kept} its community identity by being in a fragment (i and ii) or a node that was {\bf dispersed} from its original place (iii). Moreover, each algorithm will have produced a number of different {\bf fragments} that could be bigger or smaller than the number of initial communities; the former indicating that an initial community was broken into many fragments and the later indicating that different community fragments were either {\bf joined} together or completely {\bf obliterated} (all their nodes were dispersed among different fragments). The fraction of nodes kept and dispersed, the proportion of fragments and communities obliterated with respect to the initial number of communities, the number of joined fragments and the resulting number of communities for the LFR, RC and OUR benchmarks produced by each algorithm can be seen in Tables \ref{tabLFRres}, \ref{tabRCres} and \ref{tabOURres}, respectively. To facilitate the understanding of this fragmentation analysis, we illustrate an example in Scheme \ref{scheme1} where 30 nodes in 6 communities are reassigned to 5 different communities by an algorithm. The tables present the statistics over many graphs of the results produced by each network by this analysis.

\begin{scheme}[h]
\hrule{}
\bmn{0.25\tw}
\bc
\ssp

\underline{Initial}

\ssp

\bt{rc}
$c_1$: & \fbox{XXXXXX} \\
$c_2$: & \fbox{OOOO} \\
$c_3$: & \fbox{$\xi$ $\xi$ $\xi$ $\xi$ $\xi$ $\xi$ } \\
$c_4$: & \fbox{VVV} \\
$c_5$: & \fbox{******} \\
$c_6$: & \fbox{$\varphi$ $\varphi$ $\varphi$ $\varphi$ $\varphi$ $\varphi$ }
\et

\ssp

\ec
\emn
\vrule{}
\bmn{0.25\tw}
\bc
\ssp

\underline{Algorithm}

\ssp

\bt{rc}
$c_1$: & \fbox{XXXX$\xi$ $\xi$ V} \\
$c_2$: &\fbox{XXV$\xi$ $\xi$ $\xi$ }\\
$c_3$: &\fbox{$\xi$ VOOO}\\
$c_4$: &\fbox{$\varphi$O**}\\
$c_5$: &\fbox{$\varphi$ $\varphi$ $\varphi$ $\varphi$****}
\et

\ssp

\ec
\emn
\vrule{}
\bmn{0.45\tw}
\bc
\bt{|l|c|}
\hline
Kept (Comms)         & 73.33 (66.67) \\
Dispersed            & 26.67 \\
Fragments (Joined)   & 116.67 (33.33) \\
Obliterated ($N_c/N_{c0}$) & 16.67 (83.33) \\
\hline
\et
\ec
\emn
\hrule{}
\caption{Schematic example on the community fragmentation analysis. Each symbol, X, O, $\xi$, V, * and $\varphi$ represents a node. The use of different symbols, mark the community to which each node belongs in the initial (benchmarked) partition, as shown in the first column above. The second column shows the algorithm assignment of each node to a community and in the third column one can see the analysis result. From the 30 nodes, 22 (73.33\%) kept their community identity: were assigned to new communities where they form fragments from their original community (four X's in the first community, another two X's in the second, three $\xi$'s in the second, three O's in the third, two *'s in the fourth and four $\varphi$'s and four *'s joined in the last) and these represent seven fragments from the six initial communities (116.67\%) that kept somehow their identities. If one thinks in terms of conserved communities, 4 out of 6 (66.67\%) where kept: the four X's in the first are more than 50\% of all X's, the same with the three O's in the third and the four $\varphi$ and four *'s in the last. The remaining nodes were dispersed among the five communities without forming a majority. There were two fragments joined together in the second community ($\xi$'s and X's) and another two joined in the last ($\varphi$ and *), hence the 33.33\% in joined that represents 2 sixths of the number of initial communities. The community formed by the V's was obliterated (16.67\%), had its nodes completely dispersed. Finally, the last number, 83.33\% next to obliterated, represents the ratio between the number of communities identified by the algorithm ($N_c=5$) and the initial number of communities ($N_{c0}=6$).} \label{scheme1}
\end{scheme}


\bgroup
\def\arraystretch{1.5}
\begin{table}
\caption{Fragmentation analysis over LFR Benchmarks. The values in the table are averages over all graphs with the correspondent $\mu$ values indicated. The explanations on the ratios evaluated can be found in the text and in the example in Scheme \ref{scheme1}.}\label{tabLFRres}
\bc
\scriptsize
\bt{c|c|ccc}
\multirow{2}{*}{Algorithm} & \multirow{2}{*}{Characteristic} & \multicolumn{3}{c}{$\mu$} \\
 \cline{3-5}
 & & $\mu=0.0$ & $\mu\le 0.5$ & $\mu> 0.5$ \\
\hline
\hline
\multirow{4}{*}{louvain} & Kept (Comms)         & $100.00^{+ 0.00}_{- 0.00}$ ($100.00^{+ 0.00}_{- 0.00}$) & $98.78^{+ 1.19}_{-10.88}$ ($99.10^{+ 0.90}_{-23.80}$) & $75.65^{+14.53}_{- 7.47}$ ($44.16^{+40.60}_{-33.95}$) \\ 
 & Dispersed            & $ 0.00^{+ 0.00}_{- 0.00}$ & $ 1.22^{+10.88}_{- 1.19}$ & $24.35^{+ 7.47}_{-14.53}$ \\ 
 & Fragments (Joined)   & $100.23^{+13.32}_{- 0.23}$ ($ 0.59^{+ 3.65}_{- 0.59}$) & $103.89^{+79.54}_{- 3.91}$ ($18.33^{+25.85}_{-14.30}$) & $230.63^{+93.92}_{-103.06}$ ($179.10^{+88.54}_{-101.58}$) \\ 
 & Obliterated ($N_c/N_{c0}$) & $ 0.00^{+ 0.00}_{- 0.00}$ ($99.63^{+ 1.91}_{- 3.85}$) & $ 0.01^{+ 6.40}_{- 0.01}$ ($85.58^{+19.10}_{-17.73}$) & $ 6.70^{+13.14}_{- 5.43}$ ($51.53^{+63.46}_{-19.36}$) \\ 
\hline
\hline
\multirow{4}{*}{CPM} & Kept (Comms)         & $96.63^{+ 3.38}_{-32.30}$ ($91.59^{+ 8.44}_{-81.98}$) & $89.15^{+ 9.81}_{-28.34}$ ($87.13^{+12.67}_{-48.81}$) & $54.73^{+23.27}_{-19.44}$ ($30.50^{+41.68}_{-25.92}$) \\ 
 & Dispersed            & $ 3.37^{+32.30}_{- 3.38}$ & $10.85^{+28.34}_{- 9.81}$ & $45.27^{+19.44}_{-23.27}$ \\ 
 & Fragments (Joined)   & $344.59^{+2623.59}_{-244.31}$ ($ 0.51^{+ 7.47}_{- 0.51}$) & $373.50^{+1153.66}_{-253.80}$ ($28.29^{+177.68}_{-26.94}$) & $467.16^{+860.97}_{-170.42}$ ($109.42^{+227.11}_{-64.54}$) \\ 
 & Obliterated ($N_c/N_{c0}$) & $ 0.00^{+ 0.00}_{- 0.00}$ ($1024.31^{+9366.48}_{-926.69}$) & $ 0.03^{+ 7.91}_{- 0.03}$ ($872.80^{+3417.78}_{-738.81}$) & $ 3.92^{+ 8.69}_{- 3.75}$ ($601.15^{+1345.65}_{-247.93}$) \\ 
\hline
\hline
\multirow{4}{*}{Infomap} & Kept (Comms)         & $99.99^{+ 0.01}_{- 0.15}$ ($100.00^{+ 0.00}_{- 0.00}$) & $99.46^{+ 0.53}_{- 3.66}$ ($99.98^{+ 0.02}_{- 1.15}$) & $97.96^{+ 1.86}_{- 8.38}$ ($99.36^{+ 0.64}_{- 7.18}$) \\ 
 & Dispersed            & $ 0.01^{+ 0.15}_{- 0.01}$ & $ 0.54^{+ 3.66}_{- 0.53}$ & $ 2.04^{+ 8.38}_{- 1.86}$ \\ 
 & Fragments (Joined)   & $100.92^{+52.53}_{- 0.92}$ ($ 0.00^{+ 0.00}_{- 0.00}$) & $115.26^{+123.97}_{-15.05}$ ($ 3.41^{+37.91}_{- 3.41}$) & $118.68^{+113.22}_{-17.97}$ ($72.24^{+27.69}_{-64.75}$) \\ 
 & Obliterated ($N_c/N_{c0}$) & $ 0.00^{+ 0.00}_{- 0.00}$ ($101.92^{+35.69}_{- 1.93}$) & $ 0.00^{+ 0.00}_{- 0.00}$ ($117.09^{+142.83}_{-17.15}$) & $ 0.03^{+ 2.39}_{- 0.03}$ ($49.00^{+79.64}_{-44.77}$) \\ 
\hline
\hline
\multirow{4}{*}{RB} & Kept (Comms)         & $100.00^{+ 0.00}_{- 0.00}$ ($100.00^{+ 0.00}_{- 0.00}$) & $97.84^{+ 1.99}_{-13.08}$ ($97.87^{+ 2.13}_{-35.87}$) & $63.78^{+19.56}_{-20.51}$ ($36.73^{+42.39}_{-31.65}$) \\ 
 & Dispersed            & $ 0.00^{+ 0.00}_{- 0.00}$ & $ 2.16^{+13.08}_{- 1.99}$ & $36.22^{+20.51}_{-19.56}$ \\ 
 & Fragments (Joined)   & $100.00^{+ 0.00}_{- 0.00}$ ($ 0.25^{+ 6.44}_{- 0.25}$) & $127.89^{+487.64}_{-26.70}$ ($10.02^{+126.14}_{- 9.70}$) & $431.89^{+530.01}_{-186.09}$ ($178.91^{+207.92}_{-108.55}$) \\ 
 & Obliterated ($N_c/N_{c0}$) & $ 0.00^{+ 0.00}_{- 0.00}$ ($99.75^{+ 0.25}_{- 6.44}$) & $ 0.02^{+ 5.90}_{- 0.02}$ ($118.79^{+384.23}_{-20.43}$) & $ 2.68^{+ 7.37}_{- 2.59}$ ($276.07^{+351.69}_{-106.05}$) \\ 
\hline
\hline
\multirow{4}{*}{RN} & Kept (Comms)         & $100.00^{+ 0.00}_{- 0.00}$ ($100.00^{+ 0.00}_{- 0.00}$) & $98.46^{+ 1.50}_{-27.16}$ ($98.98^{+ 1.02}_{-44.34}$) & $55.71^{+39.73}_{-33.72}$ ($44.38^{+53.52}_{-43.20}$) \\ 
 & Dispersed            & $ 0.00^{+ 0.00}_{- 0.00}$ & $ 1.54^{+27.16}_{- 1.50}$ & $44.29^{+33.72}_{-39.73}$ \\ 
 & Fragments (Joined)   & $100.00^{+ 0.00}_{- 0.00}$ ($ 0.21^{+ 7.52}_{- 0.21}$) & $122.34^{+570.06}_{-22.12}$ ($12.14^{+64.18}_{-11.97}$) & $201.58^{+185.74}_{-81.33}$ ($33.59^{+94.16}_{-32.39}$) \\ 
 & Obliterated ($N_c/N_{c0}$) & $ 0.00^{+ 0.00}_{- 0.00}$ ($99.79^{+ 0.21}_{- 7.52}$) & $ 0.03^{+24.54}_{- 0.03}$ ($199.23^{+2757.73}_{-111.07}$) & $ 8.85^{+19.49}_{- 8.03}$ ($716.43^{+661.18}_{-646.13}$) \\ 
\hline
\hline
\multirow{4}{*}{SCluster} & Kept (Comms)         & $97.22^{+ 2.77}_{-15.36}$ ($93.04^{+ 6.96}_{-80.41}$) & $87.06^{+ 9.44}_{-17.37}$ ($85.95^{+13.35}_{-44.98}$) & $51.72^{+16.25}_{-17.85}$ ($21.70^{+29.77}_{-19.00}$) \\ 
 & Dispersed            & $ 2.78^{+15.36}_{- 2.77}$ & $12.94^{+17.37}_{- 9.44}$ & $48.28^{+17.85}_{-16.25}$ \\ 
 & Fragments (Joined)   & $336.59^{+2319.84}_{-231.36}$ ($ 0.02^{+ 0.67}_{- 0.02}$) & $411.53^{+1142.48}_{-276.76}$ ($47.70^{+220.67}_{-44.33}$) & $521.19^{+1030.06}_{-192.20}$ ($121.79^{+302.94}_{-69.94}$) \\ 
 & Obliterated ($N_c/N_{c0}$) & $ 0.00^{+ 0.00}_{- 0.00}$ ($722.28^{+4640.96}_{-611.37}$) & $ 0.00^{+ 0.00}_{- 0.00}$ ($675.06^{+2221.66}_{-520.03}$) & $ 3.32^{+10.19}_{- 3.25}$ ($569.88^{+1151.36}_{-188.52}$) \\ 
\hline
\hline
\multirow{4}{*}{UVCluster} & Kept (Comms)         & $98.19^{+ 1.74}_{-17.67}$ ($98.55^{+ 1.46}_{-77.32}$) & $83.55^{+10.42}_{-13.92}$ ($84.09^{+14.74}_{-40.48}$) & $48.81^{+17.18}_{-20.26}$ ($17.20^{+25.87}_{-15.27}$) \\ 
 & Dispersed            & $ 1.81^{+17.67}_{- 1.74}$ & $16.45^{+13.92}_{-10.42}$ & $51.19^{+20.26}_{-17.18}$ \\ 
 & Fragments (Joined)   & $125.77^{+626.67}_{-25.76}$ ($ 0.95^{+ 7.90}_{- 0.95}$) & $357.08^{+984.10}_{-217.21}$ ($64.24^{+268.71}_{-55.16}$) & $524.29^{+1084.76}_{-206.35}$ ($115.64^{+433.20}_{-81.61}$) \\ 
 & Obliterated ($N_c/N_{c0}$) & $ 0.00^{+ 0.00}_{- 0.00}$ ($365.71^{+3214.35}_{-266.68}$) & $ 0.00^{+ 0.00}_{- 0.00}$ ($533.63^{+1734.11}_{-383.58}$) & $ 4.50^{+12.48}_{- 4.28}$ ($606.19^{+772.85}_{-172.27}$) \\ 
\hline
\hline
\multirow{4}{*}{surpriser} & Kept (Comms)         & $91.37^{+ 8.61}_{-21.15}$ ($91.13^{+ 8.70}_{-73.77}$) & $88.32^{+10.67}_{-25.98}$ ($89.32^{+10.43}_{-45.78}$) & $57.69^{+22.80}_{-17.76}$ ($31.99^{+41.38}_{-27.65}$) \\ 
 & Dispersed            & $ 8.63^{+21.15}_{- 8.61}$ & $11.68^{+25.98}_{-10.67}$ & $42.31^{+17.76}_{-22.80}$ \\ 
 & Fragments (Joined)   & $315.16^{+1428.53}_{-204.85}$ ($ 0.00^{+ 0.00}_{- 0.00}$) & $312.64^{+873.74}_{-193.00}$ ($28.28^{+180.17}_{-27.16}$) & $484.00^{+821.93}_{-170.16}$ ($132.66^{+201.51}_{-77.04}$) \\ 
 & Obliterated ($N_c/N_{c0}$) & $ 0.00^{+ 0.00}_{- 0.00}$ ($1575.20^{+6108.83}_{-1405.25}$) & $ 0.05^{+ 6.53}_{- 0.05}$ ($972.69^{+3248.12}_{-834.46}$) & $ 3.10^{+ 8.56}_{- 3.02}$ ($558.95^{+1532.65}_{-246.10}$) \\ 
\hline
\hline
\et
\ec
\end{table}
\egroup


\bgroup
\def\arraystretch{1.5}
\begin{table}
\caption{Fragmentation analysis over RC Benchmarks. The values in the table are averages over all graphs with the correspondent $\mu$ values indicated. The explanations on the ratios evaluated can be found in the text and in the example in Scheme \ref{scheme1}.}\label{tabRCres}
\bc
\scriptsize
\bt{c|c|ccc}
\multirow{2}{*}{Algorithm} & \multirow{2}{*}{Characteristic} & \multicolumn{3}{c}{$\mu$} \\
 \cline{3-5}
 & & $\mu=0.0$ & $\mu\le 0.5$ & $\mu> 0.5$ \\
\hline
\hline
\multirow{4}{*}{louvain} & Kept (Comms)         & $100.00^{+ 0.00}_{- 0.00}$ ($100.00^{+ 0.00}_{- 0.00}$) & $97.49^{+ 2.09}_{- 6.38}$ ($93.16^{+ 5.87}_{-14.14}$) & $70.73^{+11.24}_{- 5.89}$ ($22.84^{+40.26}_{-18.45}$) \\ 
 & Dispersed            & $ 0.00^{+ 0.00}_{- 0.00}$ & $ 2.51^{+ 6.38}_{- 2.09}$ & $29.27^{+ 5.89}_{-11.24}$ \\ 
 & Fragments (Joined)   & $100.00^{+ 0.00}_{- 0.00}$ ($ 0.00^{+ 0.00}_{- 0.00}$) & $93.36^{+ 5.68}_{-13.53}$ ($46.31^{+30.29}_{-29.62}$) & $202.29^{+147.69}_{-88.98}$ ($162.74^{+134.06}_{-78.10}$) \\ 
 & Obliterated ($N_c/N_{c0}$) & $ 0.00^{+ 0.00}_{- 0.00}$ ($100.00^{+ 0.00}_{- 0.00}$) & $ 6.69^{+13.62}_{- 5.73}$ ($47.06^{+36.18}_{-27.11}$) & $32.23^{+16.39}_{-17.12}$ ($49.35^{+62.41}_{-23.93}$) \\ 
\hline
\hline
\multirow{4}{*}{CPM} & Kept (Comms)         & $100.00^{+ 0.00}_{- 0.00}$ ($100.00^{+ 0.00}_{- 0.00}$) & $97.66^{+ 1.99}_{- 5.69}$ ($89.15^{+ 9.13}_{-27.02}$) & $55.93^{+20.29}_{-15.25}$ ($ 8.75^{+27.46}_{- 8.24}$) \\ 
 & Dispersed            & $ 0.00^{+ 0.00}_{- 0.00}$ & $ 2.34^{+ 5.69}_{- 1.99}$ & $44.07^{+15.25}_{-20.29}$ \\ 
 & Fragments (Joined)   & $100.00^{+ 0.00}_{- 0.00}$ ($ 0.00^{+ 0.00}_{- 0.00}$) & $106.11^{+26.49}_{- 7.30}$ ($ 4.53^{+18.24}_{- 4.38}$) & $351.80^{+88.44}_{-130.20}$ ($89.29^{+48.34}_{-34.92}$) \\ 
 & Obliterated ($N_c/N_{c0}$) & $ 0.00^{+ 0.00}_{- 0.00}$ ($100.00^{+ 0.00}_{- 0.00}$) & $ 4.47^{+ 8.79}_{- 4.21}$ ($108.42^{+16.28}_{- 7.62}$) & $28.31^{+16.70}_{-14.88}$ ($357.52^{+138.57}_{-212.62}$) \\ 
\hline
\hline
\multirow{4}{*}{Infomap} & Kept (Comms)         & $100.00^{+ 0.00}_{- 0.00}$ ($100.00^{+ 0.00}_{- 0.00}$) & $98.21^{+ 1.52}_{- 4.34}$ ($93.09^{+ 6.05}_{-14.75}$) & $75.14^{+21.48}_{-22.61}$ ($45.92^{+48.34}_{-42.00}$) \\ 
 & Dispersed            & $ 0.00^{+ 0.00}_{- 0.00}$ & $ 1.79^{+ 4.34}_{- 1.52}$ & $24.86^{+22.61}_{-21.48}$ \\ 
 & Fragments (Joined)   & $100.00^{+ 0.00}_{- 0.00}$ ($ 0.00^{+ 0.00}_{- 0.00}$) & $94.82^{+ 4.78}_{-12.82}$ ($12.32^{+42.78}_{-10.15}$) & $257.23^{+176.46}_{-154.44}$ ($149.59^{+98.38}_{-76.00}$) \\ 
 & Obliterated ($N_c/N_{c0}$) & $ 0.00^{+ 0.00}_{- 0.00}$ ($100.00^{+ 0.00}_{- 0.00}$) & $ 5.75^{+12.94}_{- 4.94}$ ($82.71^{+14.34}_{-38.39}$) & $14.76^{+19.85}_{-12.48}$ ($124.56^{+136.19}_{-103.90}$) \\ 
\hline
\hline
\multirow{4}{*}{RB} & Kept (Comms)         & $100.00^{+ 0.00}_{- 0.00}$ ($100.00^{+ 0.00}_{- 0.00}$) & $97.83^{+ 1.82}_{- 5.72}$ ($93.19^{+ 5.89}_{-17.62}$) & $59.53^{+16.72}_{-14.34}$ ($12.36^{+42.45}_{-11.59}$) \\ 
 & Dispersed            & $ 0.00^{+ 0.00}_{- 0.00}$ & $ 2.17^{+ 5.72}_{- 1.82}$ & $40.47^{+14.34}_{-16.72}$ \\ 
 & Fragments (Joined)   & $100.00^{+ 0.00}_{- 0.00}$ ($ 0.00^{+ 0.00}_{- 0.00}$) & $96.65^{+25.54}_{-11.54}$ ($34.34^{+35.69}_{-27.67}$) & $378.92^{+92.90}_{-213.51}$ ($104.28^{+47.36}_{-40.17}$) \\ 
 & Obliterated ($N_c/N_{c0}$) & $ 0.00^{+ 0.00}_{- 0.00}$ ($100.00^{+ 0.00}_{- 0.00}$) & $ 5.81^{+13.44}_{- 4.94}$ ($62.49^{+40.67}_{-35.02}$) & $23.78^{+12.54}_{-14.03}$ ($332.54^{+120.40}_{-220.54}$) \\ 
\hline
\hline
\multirow{4}{*}{RN} & Kept (Comms)         & $100.00^{+ 0.00}_{- 0.00}$ ($100.00^{+ 0.00}_{- 0.00}$) & $99.53^{+ 0.37}_{- 1.10}$ ($98.17^{+ 1.83}_{- 5.94}$) & $92.04^{+ 7.07}_{-80.56}$ ($91.64^{+ 8.06}_{-87.79}$) \\ 
 & Dispersed            & $ 0.00^{+ 0.00}_{- 0.00}$ & $ 0.47^{+ 1.10}_{- 0.37}$ & $ 7.96^{+80.56}_{- 7.07}$ \\ 
 & Fragments (Joined)   & $100.00^{+ 0.00}_{- 0.00}$ ($ 0.00^{+ 0.00}_{- 0.00}$) & $99.47^{+ 2.55}_{- 5.09}$ ($37.94^{+56.83}_{-37.42}$) & $103.39^{+29.51}_{- 3.87}$ ($88.26^{+ 7.91}_{-84.98}$) \\ 
 & Obliterated ($N_c/N_{c0}$) & $ 0.00^{+ 0.00}_{- 0.00}$ ($100.00^{+ 0.00}_{- 0.00}$) & $ 1.44^{+ 4.80}_{- 1.44}$ ($68.16^{+34.46}_{-52.76}$) & $ 3.76^{+29.98}_{- 3.68}$ ($141.71^{+1322.24}_{-119.43}$) \\ 
\hline
\hline
\multirow{4}{*}{SCluster} & Kept (Comms)         & $97.66^{+ 2.34}_{-41.64}$ ($95.86^{+ 4.15}_{-72.57}$) & $95.90^{+ 3.47}_{-10.38}$ ($86.23^{+11.81}_{-30.64}$) & $52.20^{+16.98}_{-16.22}$ ($ 5.97^{+21.25}_{- 5.58}$) \\ 
 & Dispersed            & $ 2.34^{+41.64}_{- 2.34}$ & $ 4.10^{+10.38}_{- 3.47}$ & $47.80^{+16.22}_{-16.98}$ \\ 
 & Fragments (Joined)   & $95.92^{+ 4.08}_{-71.49}$ ($ 2.64^{+ 2.37}_{- 1.47}$) & $109.76^{+37.70}_{-14.26}$ ($ 4.73^{+14.41}_{- 4.31}$) & $361.03^{+77.77}_{-104.34}$ ($56.25^{+34.96}_{-19.74}$) \\ 
 & Obliterated ($N_c/N_{c0}$) & $ 4.08^{+71.49}_{- 4.08}$ ($151.81^{+972.48}_{-54.68}$) & $ 5.68^{+16.82}_{- 4.85}$ ($128.70^{+213.91}_{-24.38}$) & $29.40^{+16.20}_{-16.22}$ ($454.47^{+119.83}_{-248.36}$) \\ 
\hline
\hline
\multirow{4}{*}{UVCluster} & Kept (Comms)         & $97.66^{+ 2.34}_{-41.64}$ ($95.86^{+ 4.15}_{-72.57}$) & $94.75^{+ 4.42}_{-11.35}$ ($85.96^{+12.10}_{-30.26}$) & $48.99^{+16.11}_{-14.82}$ ($ 3.20^{+16.35}_{- 3.12}$) \\ 
 & Dispersed            & $ 2.34^{+41.64}_{- 2.34}$ & $ 5.25^{+11.35}_{- 4.42}$ & $51.01^{+14.82}_{-16.11}$ \\ 
 & Fragments (Joined)   & $95.92^{+ 4.08}_{-71.49}$ ($ 2.64^{+ 2.37}_{- 1.47}$) & $110.04^{+44.86}_{-14.65}$ ($10.41^{+34.69}_{- 8.65}$) & $392.86^{+73.95}_{-98.94}$ ($49.75^{+46.26}_{-21.21}$) \\ 
 & Obliterated ($N_c/N_{c0}$) & $ 4.08^{+71.49}_{- 4.08}$ ($151.81^{+972.48}_{-54.68}$) & $ 6.44^{+16.80}_{- 5.58}$ ($113.23^{+221.57}_{-16.22}$) & $30.96^{+17.39}_{-16.33}$ ($534.30^{+108.65}_{-276.87}$) \\ 
\hline
\hline
\multirow{4}{*}{surpriser} & Kept (Comms)         & $100.00^{+ 0.00}_{- 0.00}$ ($100.00^{+ 0.00}_{- 0.00}$) & $97.78^{+ 1.86}_{- 5.46}$ ($89.62^{+ 8.68}_{-27.17}$) & $55.41^{+18.13}_{-14.65}$ ($ 7.26^{+29.70}_{- 6.83}$) \\ 
 & Dispersed            & $ 0.00^{+ 0.00}_{- 0.00}$ & $ 2.22^{+ 5.46}_{- 1.86}$ & $44.59^{+14.65}_{-18.13}$ \\ 
 & Fragments (Joined)   & $100.00^{+ 0.00}_{- 0.00}$ ($ 0.00^{+ 0.00}_{- 0.00}$) & $106.46^{+27.95}_{- 7.69}$ ($ 5.12^{+23.94}_{- 4.69}$) & $377.10^{+74.10}_{-120.62}$ ($94.44^{+36.22}_{-35.98}$) \\ 
 & Obliterated ($N_c/N_{c0}$) & $ 0.00^{+ 0.00}_{- 0.00}$ ($100.00^{+ 0.00}_{- 0.00}$) & $ 4.28^{+ 8.44}_{- 4.02}$ ($109.38^{+15.01}_{- 7.99}$) & $27.57^{+16.08}_{-15.24}$ ($355.06^{+96.60}_{-152.45}$) \\ 
\hline
\hline
\et
\ec
\end{table}
\egroup


\bgroup
\def\arraystretch{1.5}
\begin{table}
\caption{Fragmentation analysis over OUR Benchmarks. The values in the table are averages over all graphs with the correspondent $\mu$ values indicated. The explanations on the ratios evaluated can be found in the text and in the example in Scheme \ref{scheme1}.}\label{tabOURres}
\bc
\scriptsize
\bt{c|c|ccc}
\multirow{2}{*}{Algorithm} & \multirow{2}{*}{Characteristic} & \multicolumn{3}{c}{$\mu$} \\
 \cline{3-5}
 & & $\mu=0.0$ & $\mu\le 0.5$ & $\mu> 0.5$ \\
\hline
\hline
\multirow{4}{*}{louvain} & Kept (Comms)         & $99.00^{+ 0.00}_{- 0.00}$ ($80.00^{+ 0.00}_{- 0.00}$) & $97.75^{+ 1.12}_{- 2.28}$ ($71.52^{+ 7.83}_{-14.87}$) & $80.32^{+10.86}_{-10.92}$ ($34.50^{+26.07}_{-26.83}$) \\ 
 & Dispersed            & $ 1.00^{+ 0.00}_{- 0.00}$ & $ 2.25^{+ 2.28}_{- 1.12}$ & $19.68^{+10.92}_{-10.86}$ \\ 
 & Fragments (Joined)   & $80.00^{+ 0.00}_{- 0.00}$ ($ 0.00^{+ 0.00}_{- 0.00}$) & $71.53^{+ 7.83}_{-14.87}$ ($19.95^{+13.14}_{-17.23}$) & $124.06^{+116.97}_{-64.26}$ ($95.65^{+114.27}_{-65.03}$) \\ 
 & Obliterated ($N_c/N_{c0}$) & $20.00^{+ 0.00}_{- 0.00}$ ($80.00^{+ 0.00}_{- 0.00}$) & $28.48^{+14.87}_{- 7.83}$ ($51.59^{+23.95}_{-19.52}$) & $40.78^{+15.39}_{-17.92}$ ($28.41^{+14.84}_{- 8.69}$) \\ 
\hline
\hline
\multirow{4}{*}{CPM} & Kept (Comms)         & $100.00^{+ 0.00}_{- 0.02}$ ($99.99^{+ 0.01}_{- 0.33}$) & $99.28^{+ 0.63}_{- 1.43}$ ($92.31^{+ 7.01}_{-14.88}$) & $71.07^{+22.22}_{-34.37}$ ($36.00^{+33.66}_{-31.87}$) \\ 
 & Dispersed            & $ 0.00^{+ 0.02}_{- 0.00}$ & $ 0.72^{+ 1.43}_{- 0.63}$ & $28.93^{+34.37}_{-22.22}$ \\ 
 & Fragments (Joined)   & $99.99^{+ 0.01}_{- 0.33}$ ($ 0.00^{+ 0.00}_{- 0.00}$) & $93.24^{+ 6.16}_{-14.05}$ ($ 0.38^{+ 3.10}_{- 0.38}$) & $175.07^{+137.15}_{-86.33}$ ($29.65^{+42.97}_{-23.40}$) \\ 
 & Obliterated ($N_c/N_{c0}$) & $ 0.01^{+ 0.33}_{- 0.01}$ ($99.99^{+ 0.01}_{- 0.33}$) & $ 7.37^{+14.33}_{- 6.70}$ ($98.72^{+ 3.84}_{- 8.81}$) & $34.34^{+17.89}_{-15.20}$ ($208.73^{+199.05}_{-115.92}$) \\ 
\hline
\hline
\multirow{4}{*}{Infomap} & Kept (Comms)         & $99.00^{+ 0.00}_{- 0.00}$ ($80.00^{+ 0.00}_{- 0.00}$) & $98.52^{+ 0.55}_{- 1.33}$ ($77.06^{+ 5.53}_{-11.64}$) & $97.73^{+ 1.25}_{- 4.08}$ ($74.18^{+ 5.80}_{-19.63}$) \\ 
 & Dispersed            & $ 1.00^{+ 0.00}_{- 0.00}$ & $ 1.48^{+ 1.33}_{- 0.55}$ & $ 2.27^{+ 4.08}_{- 1.25}$ \\ 
 & Fragments (Joined)   & $80.00^{+ 0.00}_{- 0.00}$ ($ 0.00^{+ 0.00}_{- 0.00}$) & $77.19^{+ 5.43}_{-11.42}$ ($ 3.40^{+ 6.24}_{- 3.33}$) & $74.53^{+ 5.45}_{-18.86}$ ($53.71^{+23.45}_{-46.61}$) \\ 
 & Obliterated ($N_c/N_{c0}$) & $20.00^{+ 0.00}_{- 0.00}$ ($80.00^{+ 0.00}_{- 0.00}$) & $22.86^{+11.48}_{- 5.47}$ ($74.60^{+ 7.22}_{-14.70}$) & $25.53^{+18.95}_{- 5.51}$ ($20.90^{+38.91}_{-18.06}$) \\ 
\hline
\hline
\multirow{4}{*}{RB} & Kept (Comms)         & $99.00^{+ 0.00}_{- 0.00}$ ($80.00^{+ 0.00}_{- 0.00}$) & $98.54^{+ 0.86}_{- 1.42}$ ($78.85^{+11.77}_{-13.39}$) & $75.69^{+18.50}_{-29.41}$ ($37.07^{+30.22}_{-33.22}$) \\ 
 & Dispersed            & $ 1.00^{+ 0.00}_{- 0.00}$ & $ 1.46^{+ 1.42}_{- 0.86}$ & $24.31^{+29.41}_{-18.50}$ \\ 
 & Fragments (Joined)   & $80.00^{+ 0.00}_{- 0.00}$ ($ 0.00^{+ 0.00}_{- 0.00}$) & $78.96^{+11.68}_{-13.31}$ ($10.27^{+14.38}_{- 9.10}$) & $178.24^{+164.86}_{-105.49}$ ($37.14^{+45.26}_{-27.12}$) \\ 
 & Obliterated ($N_c/N_{c0}$) & $20.00^{+ 0.00}_{- 0.00}$ ($80.00^{+ 0.00}_{- 0.00}$) & $21.08^{+13.32}_{-11.71}$ ($69.17^{+20.35}_{-21.94}$) & $35.66^{+14.86}_{-13.80}$ ($175.24^{+187.47}_{-113.36}$) \\ 
\hline
\hline
\multirow{4}{*}{RN} & Kept (Comms)         & $99.99^{+ 0.01}_{- 0.16}$ ($99.71^{+ 0.29}_{- 3.16}$) & $99.37^{+ 0.56}_{- 1.05}$ ($91.84^{+ 7.12}_{-13.76}$) & $96.77^{+ 1.81}_{-36.59}$ ($77.80^{+ 4.87}_{-22.02}$) \\ 
 & Dispersed            & $ 0.01^{+ 0.16}_{- 0.01}$ & $ 0.63^{+ 1.05}_{- 0.56}$ & $ 3.23^{+36.59}_{- 1.81}$ \\ 
 & Fragments (Joined)   & $99.71^{+ 0.29}_{- 3.16}$ ($ 0.00^{+ 0.00}_{- 0.00}$) & $92.18^{+ 7.18}_{-12.58}$ ($12.37^{+60.13}_{-11.93}$) & $82.61^{+38.30}_{- 5.72}$ ($64.57^{+11.32}_{-55.86}$) \\ 
 & Obliterated ($N_c/N_{c0}$) & $ 0.29^{+ 3.16}_{- 0.29}$ ($99.71^{+ 0.29}_{- 3.16}$) & $ 8.05^{+13.66}_{- 7.00}$ ($83.92^{+16.33}_{-64.81}$) & $20.92^{+ 9.96}_{- 4.29}$ ($37.04^{+255.54}_{-25.74}$) \\ 
\hline
\hline
\multirow{4}{*}{SCluster} & Kept (Comms)         & $99.89^{+ 0.08}_{- 0.13}$ ($97.81^{+ 1.52}_{- 2.59}$) & $99.09^{+ 0.74}_{- 1.60}$ ($91.47^{+ 6.70}_{-15.71}$) & $68.13^{+20.67}_{-29.05}$ ($31.89^{+32.52}_{-28.32}$) \\ 
 & Dispersed            & $ 0.11^{+ 0.13}_{- 0.08}$ & $ 0.91^{+ 1.60}_{- 0.74}$ & $31.87^{+29.05}_{-20.67}$ \\ 
 & Fragments (Joined)   & $97.81^{+ 1.52}_{- 2.59}$ ($ 0.00^{+ 0.00}_{- 0.00}$) & $92.48^{+ 6.32}_{-13.14}$ ($ 0.34^{+ 3.08}_{- 0.34}$) & $191.59^{+120.30}_{-92.64}$ ($31.57^{+36.28}_{-25.12}$) \\ 
 & Obliterated ($N_c/N_{c0}$) & $ 2.19^{+ 2.59}_{- 1.52}$ ($97.81^{+ 1.52}_{- 2.59}$) & $ 8.19^{+15.16}_{- 6.39}$ ($98.76^{+ 8.09}_{- 6.92}$) & $35.48^{+17.56}_{-15.16}$ ($226.51^{+149.75}_{-108.15}$) \\ 
\hline
\hline
\multirow{4}{*}{UVCluster} & Kept (Comms)         & $99.89^{+ 0.08}_{- 0.13}$ ($97.81^{+ 1.52}_{- 2.59}$) & $98.57^{+ 1.05}_{- 2.12}$ ($86.55^{+ 9.05}_{-15.33}$) & $63.82^{+21.78}_{-28.06}$ ($24.99^{+33.71}_{-23.38}$) \\ 
 & Dispersed            & $ 0.11^{+ 0.13}_{- 0.08}$ & $ 1.43^{+ 2.12}_{- 1.05}$ & $36.18^{+28.06}_{-21.78}$ \\ 
 & Fragments (Joined)   & $97.81^{+ 1.52}_{- 2.59}$ ($ 0.00^{+ 0.00}_{- 0.00}$) & $87.69^{+ 7.91}_{-14.74}$ ($ 0.89^{+ 4.01}_{- 0.89}$) & $211.95^{+114.10}_{-112.22}$ ($26.15^{+23.29}_{-16.38}$) \\ 
 & Obliterated ($N_c/N_{c0}$) & $ 2.19^{+ 2.59}_{- 1.52}$ ($97.81^{+ 1.52}_{- 2.59}$) & $13.13^{+14.83}_{- 8.76}$ ($91.46^{+ 6.33}_{-12.12}$) & $38.17^{+17.14}_{-15.01}$ ($272.71^{+162.17}_{-172.58}$) \\ 
\hline
\hline
\multirow{4}{*}{surpriser} & Kept (Comms)         & $99.99^{+ 0.01}_{- 0.14}$ ($99.71^{+ 0.29}_{- 2.72}$) & $99.28^{+ 0.62}_{- 1.40}$ ($92.20^{+ 6.92}_{-14.45}$) & $70.24^{+23.22}_{-29.37}$ ($33.52^{+36.90}_{-31.49}$) \\ 
 & Dispersed            & $ 0.01^{+ 0.14}_{- 0.01}$ & $ 0.72^{+ 1.40}_{- 0.62}$ & $29.76^{+29.37}_{-23.22}$ \\ 
 & Fragments (Joined)   & $99.71^{+ 0.29}_{- 2.72}$ ($ 0.00^{+ 0.00}_{- 0.00}$) & $92.92^{+ 6.27}_{-13.76}$ ($ 0.42^{+ 2.97}_{- 0.42}$) & $201.96^{+125.72}_{-111.10}$ ($42.05^{+46.00}_{-34.50}$) \\ 
 & Obliterated ($N_c/N_{c0}$) & $ 0.29^{+ 2.72}_{- 0.29}$ ($99.71^{+ 0.29}_{- 2.72}$) & $ 7.54^{+14.03}_{- 6.67}$ ($98.45^{+ 4.16}_{- 7.81}$) & $33.99^{+18.33}_{-15.92}$ ($202.69^{+116.45}_{-106.46}$) \\ 
\hline
\hline
\et
\ec
\end{table}
\egroup

Analysing the information in Tables \ref{tabLFRres}, \ref{tabRCres} and \ref{tabOURres}, it is possible to understand what is the discrepancy (if any) found by a given algorithm in a given benchmark. Reading the column $\mu=0.0$ in Table \ref{tabLFRres}, for example, one sees that the louvain algorithm over LFR graphs keeps 100\% of a network's nodes mostly in fragments (communities) with other nodes from the initial benchmark community assignment and the fact that usually the number of fragments produced by this algorithm is close (100.23\%) to the initial number of communities is telling us that this algorithm in this benchmark is correctly assigning the same number of communities and in each community one finds the same nodes as the initial partition, only in a few cases a small fragment of a given community is joined to another one (0.59\%). Indeed, the value of $N_c/N_{c0}$ in this case is 99.63\%, indicating that also the number of communities identified by the algorithm is almost equal to the number of communities in the benchmarked partition. On the other hand, the surpriser algorithm is keeping around 91.37\% of the nodes in fragments with other nodes from the original community and in 91.13\% of the communities, a fragment with more than 50\% of the original community nodes is found. The reason for the surpriser average VI being bigger than the louvain's in this benchmark (the partition usually produced by the surpriser algorithm is more discrepant to the initial one than the one produced by the louvain algorithm) can be seen in the number of fragments one finds in the surpriser partition, around 3.15 (315.16\%) times the number of initial communities. Moreover, surpriser is identifying almost 16 times (1575.2\%) more communities than there were in the initial partition, many of those with just one node, such that they don't count as fragments. What the surpriser algorithm is doing is splitting some of the initial communities in many pieces. The same happens with the CPM and SCluster algorithms. The reason for this splitting can be traced back to the way in which the benchmark produces its communities: it generates independently the degree for the nodes in the community and the community size such that one can have different groups of nodes which are densely connected among them but poorly connected to each other. So, communities generated by the LFR benchmark end up having a strong subcommunity structure, though the benchmark partition itself does not notice it and algorithms with a strong resolution resolve these structures as different communities. The algorithms that work better in this benchmark are Louvain, Infomap, RB, RN and a bit behind those, UVCluster. These identify most nodes in fragments keeping their community identity (kept $\sim$ 100\%) and most communities are conserved in the algorithm partition (Comm $\sim$100\%). Actually, all algorithms perform well in this sense, but looking at the fragments, only these 5 have a number of fragments similar to the number of initial communities (fragments $\sim$100\%  and $N_c/N_{c0} \sim$100\%), others tend to divide the communities into many more pieces than in the original partition, an issue we associate to the better resolution of these other algorithms and a potential glitch of the LFR benchmark, which uses a local and not a global criterion to set its communities.

Now, for $\mu=0$, the RC benchmark produces graphs composed of disconnected cliques and nodes belonging to each separated clique are labeled as belonging to a different community. This might be the strongest definition of a community, the densest possible set of interconnected nodes with no connection whatsoever outside a given group. One would think that resolving such structures should be trivial, but focusing in Table \ref{tabRCres}, two algorithms, SCluster and UVCluster fail the task in a few cases. SCluster and UVCluster split some cliques in different fragments and some times merge different cliques together. Moreover, in this case (communities formed by cliques), they give the exact same results. These two algorithms are similar, but remembering their results for the LFR benchmark, commented above, one sees that one difference between the two is that SCluster has finer resolution than UVCluster. The reason these algorithms sometimes fail to correctly resolve these sets of unconnected cliques is, we believe, related to the rooting of the tree needed to construct the dendrogram used to identify the communities by the algorithm. 

Finally, in Table \ref{tabOURres}, one can observe the fragmentation analysis over the OUR type of benchmarks. For $\mu=0.0$, the benchmarked partition is very similar to the ones for the RC benchmark, the only difference in this case would be the single node communities added at random to the graphs. In terms of topology these communities are a single node with a link to a node from one of the cliques found in the graph. One clearly sees here that the algorithms identified with worse resolution (louvain, RB and Infomap) for their performance in the LFR benchmarks precisely miss these single nodes communities, merging the corresponding nodes to the cliques to which they are attached (in the table one sees that exactly 99\% of the nodes are kept, along with 80\% of the communities and the missing 20\% of communities are obliterated\footnote{The networks were generated with $r=0.01$, which means that 1\% of the network nodes were assigned to their own communities alone. In the small set, that means 5 communities out of 25 and in the big set 10 communities out of 50 (20\% in both cases).}). The other algorithms do not perform perfectly, but closer to the 100\% nodes and communities kept. The reason for the small discrepancy ($<1$\%) is that when a single node is attached to a size 2 clique (two nodes connected to each other), the partition with highest value for the surprise becomes degenerated. Such a 3-path subgraph can be partitioned in two different ways resulting in the same topology and from the point of view of the benchmark, the node assignment to communities in such a situation seems arbitrary. In the next subsection we comment again on this issue when analysing an specific graph. 

From Tables \ref{tabLFRres}, \ref{tabRCres} and \ref{tabOURres} one can also observe how algorithms behave as the degradation process proceeds (as $\mu$ increases). In all of them, as $\mu$ increases, more nodes are dispersed, but in some algorithms this effect is much smoother than in others, some times even not noticeable because of other effects. For example, if many communities are joined, nodes from these communities are keeping their identity and also the communities as a whole, though they are now mixed together. Therefore algorithms that tend to join communities together, mostly keep the node's identity through this fragmentation analysis. This is the case for Infomap mainly in LFR and OUR benchmarks and most notorious for RN in OUR benchmark. Actually, these two algorithms, for values of $\mu$ bigger than 0.6 tend to join all nodes in a single community, while louvain is a middle term and other algorithms usually fragment big communities in many small ones in these cases of randomly mixed connections.

Something that could be worthwhile to comment on is that some of the issues discussed above can be understood by the biases the different algorithms have with respect to the pielou index (PI) corresponding to the communities sizes. As the value of the pielou index increases so does the performance of the modularity based algorithms (Louvain and RB). This happens because for higher values of PI the community sizes become more even (for reference a PI of one means that every community has the same size) and so the algorithms can find a resolution that balances splitting larger communities into smaller ones and merging small communities into bigger ones. On the other hand for small values of PI, where we have very large communities co-existing with very small ones, those algorithms fail to find this balance and as such have a worse performance. Such algorithms tend to divide the graphs in groups of similar sizes. We could also comment that, from the 5 best performing algorithms in the LFR benchmark, Infomap and UVCluster also seem to have the same behavior as Louvain and RB which could suggest that they also suffer from a resolution limit of some sort.

To further remark on this point, we produced tables \ref{tabRCpielou} and \ref{tabOURpielou} that show, for the RC and OUR benchmarks (those where one can control the pielou index for the initial communities sizes), the evolution of the algorithm identified communities sizes pielou index as one degrades the original structure. One can see that Louvain and RB even for small degradation quite fast return partitions with pielou index close to 1, meaning even sized communities. Other algorithms, like Infomap and RN, tend to decrease the pielou index of their partitions, this happens because these algorithms tend to join nodes forming a big community sometimes encompassing all network nodes (partition for which the pielou index would be zero). Only CPM and surpriser keep the correct pielou index till a $\mu$ value arround 0.6. An interesting feature of the Infomap algorithm is that although as $\mu$ increases it tends to join all nodes in the same community ($PI\sim0$), one observes for the RC benchmark that there is a critical density point in the graphs for which it stars breaking this big community into smaller ones again, since for $\mu=0.9$ one can see in Table \ref{tabRCpielou} that the pielou index jumps again to values close to 1. In the way the RC benchmark is degraded, at $\mu=0.9$, 90\% of its links have been lost, therefore the connection density in the graph becomes strongly diminished sometimes creating many disconnected components in the graph. To visualize this issue, we show figure \ref{infomap}, where we plot the number of communities identified by the Infomap algorithm against the graph's density (number of actual links divided by the number of possible links) for each RC graph in each set at different degradation levels. One observes that, below a certain density, all nodes are merged into a single community, but as the density becomes even smaller, the graph becomes fragmented and again many communities are identified.

\bfig
\bc
\bt{cc}
\ig{0.45}{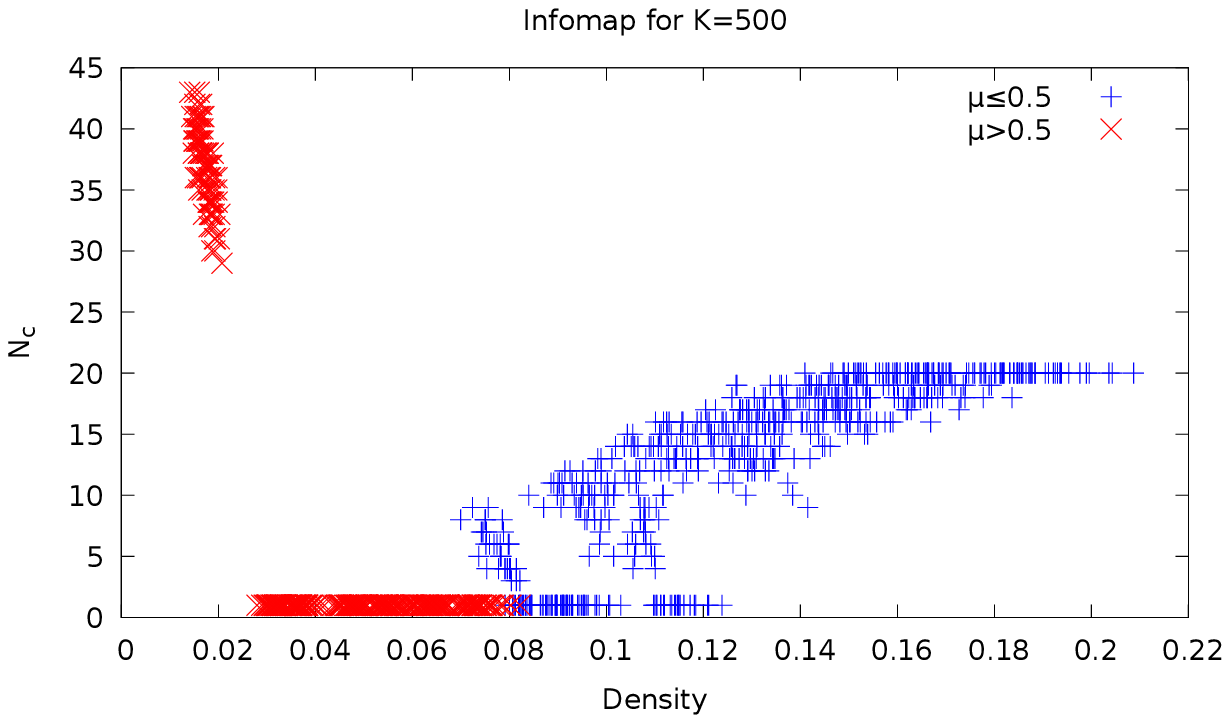} & \ig{0.45}{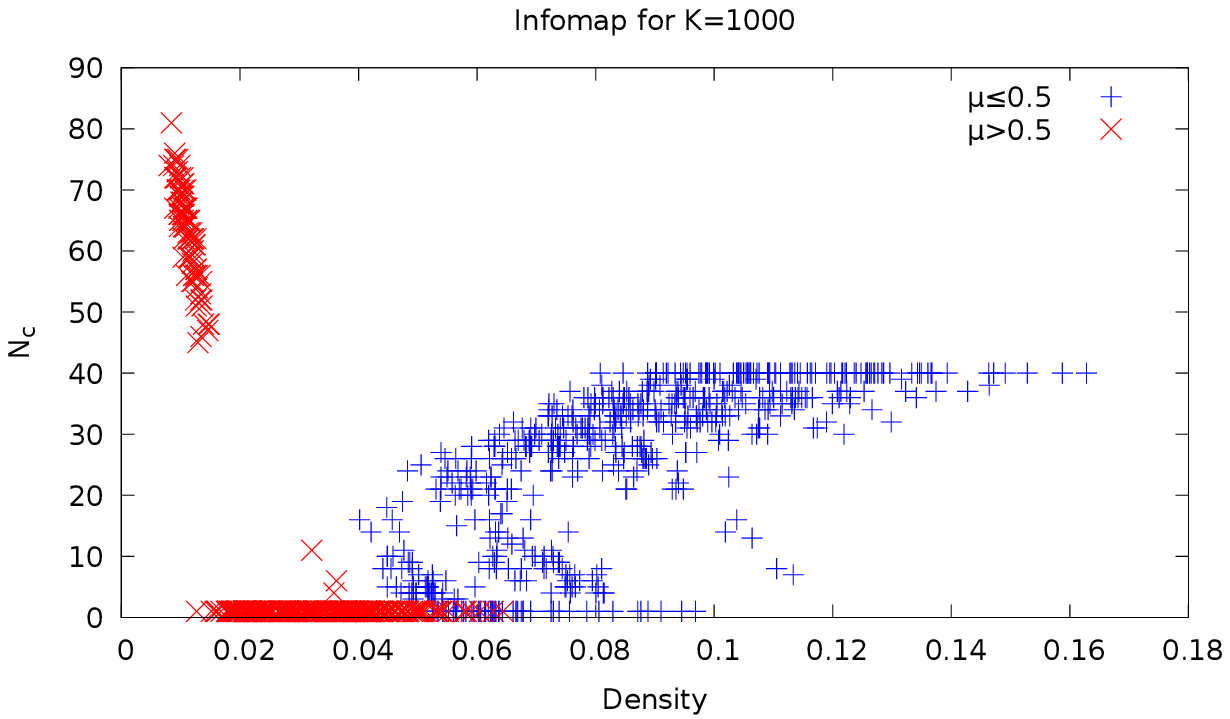}
\et
\ec
\caption{Number of communities identified by the Infomap algorithm versus graph density. Each point represents an RC benchmark at different stage of degradation (different $\mu$). We differentiate those for which $\mu$ is either bigger or smaller than 0.5.} \label{infomap}
\efig


\bgroup
\def\arraystretch{2.1}
\btab
\caption{Pielou index for the communities sizes identified by each algorithm for the RC Benchmarks. The values in the table are averages over all 100 graphs for each value of $\mu$ and benchmarked (generated) pielou index.}\label{tabRCpielou}
\bc
\tiny
\bt{c||c|cccccccccc}
\multirow{2}{*}{Algorithm} & \multirow{2}{*}{Pielou} & \multicolumn{10}{c}{$\mu$} \\
 \cline{3-12}
 & & 0.0 & 0.1 & 0.2 & 0.3 & 0.4 & 0.5 & 0.6 & 0.7 & 0.8 & 0.9 \\
\hline
\hline
\multirow{3}{*}{louvain}  & 0.75 & $ 0.76^{+ 0.02}_{- 0.01}$ & $ 0.88^{+ 0.06}_{- 0.06}$ & $ 0.90^{+ 0.05}_{- 0.07}$ & $ 0.91^{+ 0.05}_{- 0.07}$ & $ 0.92^{+ 0.05}_{- 0.07}$ & $ 0.93^{+ 0.04}_{- 0.07}$ & $ 0.93^{+ 0.05}_{- 0.09}$ & $ 0.96^{+ 0.02}_{- 0.03}$ & $ 0.96^{+ 0.01}_{- 0.02}$ & $ 0.97^{+ 0.01}_{- 0.03}$ \\ 
\cline{2-12}
    & 0.85 & $ 0.85^{+ 0.00}_{- 0.00}$ & $ 0.95^{+ 0.02}_{- 0.02}$ & $ 0.94^{+ 0.03}_{- 0.05}$ & $ 0.94^{+ 0.03}_{- 0.06}$ & $ 0.95^{+ 0.03}_{- 0.06}$ & $ 0.96^{+ 0.02}_{- 0.05}$ & $ 0.97^{+ 0.02}_{- 0.03}$ & $ 0.96^{+ 0.02}_{- 0.03}$ & $ 0.95^{+ 0.02}_{- 0.03}$ & $ 0.93^{+ 0.03}_{- 0.03}$ \\ 
\cline{2-12}
    & 0.95 & $ 0.95^{+ 0.00}_{- 0.00}$ & $ 0.98^{+ 0.01}_{- 0.01}$ & $ 0.98^{+ 0.01}_{- 0.01}$ & $ 0.98^{+ 0.00}_{- 0.01}$ & $ 0.98^{+ 0.01}_{- 0.01}$ & $ 0.98^{+ 0.01}_{- 0.01}$ & $ 0.97^{+ 0.01}_{- 0.02}$ & $ 0.95^{+ 0.02}_{- 0.02}$ & $ 0.95^{+ 0.02}_{- 0.03}$ & $ 0.79^{+ 0.03}_{- 0.02}$ \\ 
\cline{2-12}
\hline
\hline
\multirow{3}{*}{CPM}  & 0.75 & $ 0.76^{+ 0.02}_{- 0.01}$ & $ 0.76^{+ 0.02}_{- 0.02}$ & $ 0.75^{+ 0.02}_{- 0.02}$ & $ 0.75^{+ 0.02}_{- 0.02}$ & $ 0.75^{+ 0.02}_{- 0.03}$ & $ 0.76^{+ 0.02}_{- 0.02}$ & $ 0.79^{+ 0.02}_{- 0.02}$ & $ 0.91^{+ 0.07}_{- 0.08}$ & $ 0.99^{+ 0.00}_{- 0.00}$ & $ 0.99^{+ 0.00}_{- 0.00}$ \\ 
\cline{2-12}
    & 0.85 & $ 0.85^{+ 0.00}_{- 0.00}$ & $ 0.84^{+ 0.01}_{- 0.01}$ & $ 0.84^{+ 0.01}_{- 0.01}$ & $ 0.83^{+ 0.01}_{- 0.02}$ & $ 0.83^{+ 0.02}_{- 0.02}$ & $ 0.83^{+ 0.02}_{- 0.02}$ & $ 0.85^{+ 0.01}_{- 0.01}$ & $ 0.95^{+ 0.04}_{- 0.05}$ & $ 0.99^{+ 0.00}_{- 0.00}$ & $ 0.99^{+ 0.00}_{- 0.00}$ \\ 
\cline{2-12}
    & 0.95 & $ 0.95^{+ 0.00}_{- 0.00}$ & $ 0.95^{+ 0.00}_{- 0.01}$ & $ 0.94^{+ 0.01}_{- 0.01}$ & $ 0.93^{+ 0.01}_{- 0.01}$ & $ 0.92^{+ 0.02}_{- 0.02}$ & $ 0.91^{+ 0.01}_{- 0.02}$ & $ 0.92^{+ 0.01}_{- 0.01}$ & $ 0.99^{+ 0.00}_{- 0.00}$ & $ 0.99^{+ 0.00}_{- 0.00}$ & $ 0.98^{+ 0.00}_{- 0.01}$ \\ 
\cline{2-12}
\hline
\hline
\multirow{3}{*}{Infomap}  & 0.75 & $ 0.76^{+ 0.02}_{- 0.01}$ & $ 0.78^{+ 0.02}_{- 0.02}$ & $ 0.80^{+ 0.02}_{- 0.02}$ & $ 0.82^{+ 0.02}_{- 0.02}$ & $ 0.71^{+ 0.13}_{- 0.72}$ & $ 0.35^{+ 0.54}_{- 0.36}$ & $ 0.01^{+ 1.08}_{- 0.01}$ & $ 0.00^{+ 0.00}_{- 0.00}$ & $ 0.00^{+ 0.00}_{- 0.00}$ & $ 0.96^{+ 0.02}_{- 0.58}$ \\ 
\cline{2-12}
    & 0.85 & $ 0.85^{+ 0.00}_{- 0.00}$ & $ 0.86^{+ 0.01}_{- 0.01}$ & $ 0.87^{+ 0.01}_{- 0.01}$ & $ 0.88^{+ 0.01}_{- 0.01}$ & $ 0.89^{+ 0.01}_{- 0.01}$ & $ 0.90^{+ 0.02}_{- 0.02}$ & $ 0.68^{+ 0.20}_{- 0.69}$ & $ 0.10^{+ 0.76}_{- 0.10}$ & $ 0.82^{+ 0.15}_{- 0.76}$ & $ 0.98^{+ 0.01}_{- 0.01}$ \\ 
\cline{2-12}
    & 0.95 & $ 0.95^{+ 0.00}_{- 0.00}$ & $ 0.95^{+ 0.00}_{- 0.00}$ & $ 0.95^{+ 0.00}_{- 0.00}$ & $ 0.95^{+ 0.01}_{- 0.00}$ & $ 0.95^{+ 0.01}_{- 0.01}$ & $ 0.95^{+ 0.01}_{- 0.01}$ & $ 0.93^{+ 0.01}_{- 0.01}$ & $ 0.95^{+ 0.01}_{- 0.01}$ & $ 0.98^{+ 0.00}_{- 0.01}$ & $ 0.96^{+ 0.01}_{- 0.01}$ \\ 
\cline{2-12}
\hline
\hline
\multirow{3}{*}{RB}  & 0.75 & $ 0.76^{+ 0.02}_{- 0.01}$ & $ 0.90^{+ 0.06}_{- 0.06}$ & $ 0.91^{+ 0.04}_{- 0.07}$ & $ 0.91^{+ 0.05}_{- 0.07}$ & $ 0.91^{+ 0.05}_{- 0.08}$ & $ 0.91^{+ 0.04}_{- 0.06}$ & $ 0.95^{+ 0.03}_{- 0.06}$ & $ 0.98^{+ 0.01}_{- 0.06}$ & $ 0.99^{+ 0.00}_{- 0.00}$ & $ 0.99^{+ 0.00}_{- 0.00}$ \\ 
\cline{2-12}
    & 0.85 & $ 0.85^{+ 0.00}_{- 0.00}$ & $ 0.94^{+ 0.01}_{- 0.01}$ & $ 0.94^{+ 0.01}_{- 0.02}$ & $ 0.94^{+ 0.01}_{- 0.03}$ & $ 0.95^{+ 0.01}_{- 0.02}$ & $ 0.95^{+ 0.01}_{- 0.02}$ & $ 0.94^{+ 0.02}_{- 0.02}$ & $ 0.98^{+ 0.00}_{- 0.06}$ & $ 0.99^{+ 0.00}_{- 0.00}$ & $ 0.99^{+ 0.00}_{- 0.00}$ \\ 
\cline{2-12}
    & 0.95 & $ 0.95^{+ 0.00}_{- 0.00}$ & $ 0.96^{+ 0.01}_{- 0.01}$ & $ 0.96^{+ 0.01}_{- 0.01}$ & $ 0.96^{+ 0.01}_{- 0.01}$ & $ 0.97^{+ 0.01}_{- 0.01}$ & $ 0.97^{+ 0.01}_{- 0.01}$ & $ 0.96^{+ 0.01}_{- 0.01}$ & $ 0.99^{+ 0.00}_{- 0.00}$ & $ 0.99^{+ 0.00}_{- 0.00}$ & $ 0.97^{+ 0.00}_{- 0.01}$ \\ 
\cline{2-12}
\hline
\hline
\multirow{3}{*}{RN}  & 0.75 & $ 0.76^{+ 0.02}_{- 0.01}$ & $ 0.73^{+ 0.04}_{- 0.58}$ & $ 0.58^{+ 0.18}_{- 0.56}$ & $ 0.17^{+ 0.59}_{- 0.14}$ & $ 0.07^{+ 0.49}_{- 0.04}$ & $ 0.04^{+ 0.09}_{- 0.02}$ & $ 0.04^{+ 0.03}_{- 0.02}$ & $ 0.04^{+ 0.03}_{- 0.02}$ & $ 0.03^{+ 0.03}_{- 0.02}$ & $ 0.02^{+ 0.02}_{- 0.01}$ \\ 
\cline{2-12}
    & 0.85 & $ 0.85^{+ 0.00}_{- 0.00}$ & $ 0.82^{+ 0.03}_{- 0.87}$ & $ 0.76^{+ 0.08}_{- 0.76}$ & $ 0.50^{+ 0.34}_{- 0.48}$ & $ 0.10^{+ 0.63}_{- 0.07}$ & $ 0.04^{+ 0.10}_{- 0.02}$ & $ 0.04^{+ 0.03}_{- 0.02}$ & $ 0.03^{+ 0.03}_{- 0.01}$ & $ 0.03^{+ 0.02}_{- 0.01}$ & $ 0.07^{+ 0.04}_{- 0.03}$ \\ 
\cline{2-12}
    & 0.95 & $ 0.95^{+ 0.00}_{- 0.00}$ & $ 0.95^{+ 0.00}_{- 0.00}$ & $ 0.94^{+ 0.01}_{- 0.30}$ & $ 0.87^{+ 0.07}_{- 0.88}$ & $ 0.41^{+ 0.53}_{- 0.39}$ & $ 0.04^{+ 0.19}_{- 0.02}$ & $ 0.03^{+ 0.03}_{- 0.01}$ & $ 0.03^{+ 0.02}_{- 0.01}$ & $ 0.05^{+ 0.04}_{- 0.02}$ & $ 0.95^{+ 0.04}_{- 0.87}$ \\ 
\cline{2-12}
\hline
\hline
\multirow{3}{*}{SCluster}  & 0.75 & $ 0.71^{+ 0.04}_{- 0.04}$ & $ 0.75^{+ 0.02}_{- 0.03}$ & $ 0.75^{+ 0.02}_{- 0.03}$ & $ 0.74^{+ 0.03}_{- 0.03}$ & $ 0.73^{+ 0.03}_{- 0.02}$ & $ 0.74^{+ 0.02}_{- 0.02}$ & $ 0.78^{+ 0.02}_{- 0.02}$ & $ 0.92^{+ 0.06}_{- 0.08}$ & $ 0.99^{+ 0.00}_{- 0.00}$ & $ 0.99^{+ 0.00}_{- 0.00}$ \\ 
\cline{2-12}
    & 0.85 & $ 0.80^{+ 0.03}_{- 0.06}$ & $ 0.84^{+ 0.01}_{- 0.03}$ & $ 0.84^{+ 0.01}_{- 0.02}$ & $ 0.83^{+ 0.01}_{- 0.02}$ & $ 0.82^{+ 0.02}_{- 0.02}$ & $ 0.82^{+ 0.01}_{- 0.01}$ & $ 0.85^{+ 0.01}_{- 0.01}$ & $ 0.98^{+ 0.00}_{- 0.05}$ & $ 0.99^{+ 0.00}_{- 0.00}$ & $ 0.99^{+ 0.00}_{- 0.00}$ \\ 
\cline{2-12}
    & 0.95 & $ 0.94^{+ 0.01}_{- 0.01}$ & $ 0.95^{+ 0.00}_{- 0.01}$ & $ 0.94^{+ 0.01}_{- 0.01}$ & $ 0.93^{+ 0.01}_{- 0.01}$ & $ 0.92^{+ 0.01}_{- 0.02}$ & $ 0.90^{+ 0.01}_{- 0.01}$ & $ 0.94^{+ 0.02}_{- 0.01}$ & $ 0.99^{+ 0.00}_{- 0.00}$ & $ 0.99^{+ 0.00}_{- 0.00}$ & $ 0.98^{+ 0.00}_{- 0.00}$ \\ 
\cline{2-12}
\hline
\hline
\multirow{3}{*}{UVCluster}  & 0.75 & $ 0.71^{+ 0.04}_{- 0.04}$ & $ 0.76^{+ 0.02}_{- 0.03}$ & $ 0.77^{+ 0.02}_{- 0.03}$ & $ 0.77^{+ 0.03}_{- 0.03}$ & $ 0.78^{+ 0.02}_{- 0.03}$ & $ 0.80^{+ 0.02}_{- 0.02}$ & $ 0.91^{+ 0.07}_{- 0.06}$ & $ 0.99^{+ 0.00}_{- 0.00}$ & $ 0.99^{+ 0.00}_{- 0.00}$ & $ 0.99^{+ 0.00}_{- 0.00}$ \\ 
\cline{2-12}
    & 0.85 & $ 0.80^{+ 0.03}_{- 0.06}$ & $ 0.85^{+ 0.01}_{- 0.01}$ & $ 0.85^{+ 0.01}_{- 0.01}$ & $ 0.85^{+ 0.02}_{- 0.02}$ & $ 0.84^{+ 0.02}_{- 0.02}$ & $ 0.85^{+ 0.01}_{- 0.01}$ & $ 0.93^{+ 0.05}_{- 0.04}$ & $ 0.99^{+ 0.00}_{- 0.00}$ & $ 0.99^{+ 0.00}_{- 0.00}$ & $ 0.99^{+ 0.00}_{- 0.00}$ \\ 
\cline{2-12}
    & 0.95 & $ 0.94^{+ 0.01}_{- 0.01}$ & $ 0.95^{+ 0.00}_{- 0.01}$ & $ 0.95^{+ 0.00}_{- 0.01}$ & $ 0.94^{+ 0.01}_{- 0.01}$ & $ 0.93^{+ 0.01}_{- 0.01}$ & $ 0.92^{+ 0.01}_{- 0.01}$ & $ 0.97^{+ 0.02}_{- 0.02}$ & $ 0.99^{+ 0.00}_{- 0.00}$ & $ 0.99^{+ 0.00}_{- 0.00}$ & $ 0.98^{+ 0.00}_{- 0.00}$ \\ 
\cline{2-12}
\hline
\hline
\multirow{3}{*}{surpriser}  & 0.75 & $ 0.76^{+ 0.02}_{- 0.01}$ & $ 0.76^{+ 0.02}_{- 0.02}$ & $ 0.75^{+ 0.02}_{- 0.02}$ & $ 0.75^{+ 0.02}_{- 0.02}$ & $ 0.75^{+ 0.02}_{- 0.02}$ & $ 0.76^{+ 0.02}_{- 0.02}$ & $ 0.88^{+ 0.09}_{- 0.08}$ & $ 0.99^{+ 0.00}_{- 0.00}$ & $ 0.99^{+ 0.00}_{- 0.00}$ & $ 0.99^{+ 0.00}_{- 0.00}$ \\ 
\cline{2-12}
    & 0.85 & $ 0.85^{+ 0.00}_{- 0.00}$ & $ 0.85^{+ 0.01}_{- 0.01}$ & $ 0.84^{+ 0.01}_{- 0.01}$ & $ 0.83^{+ 0.01}_{- 0.02}$ & $ 0.82^{+ 0.01}_{- 0.02}$ & $ 0.83^{+ 0.01}_{- 0.02}$ & $ 0.87^{+ 0.06}_{- 0.02}$ & $ 0.98^{+ 0.01}_{- 0.04}$ & $ 0.99^{+ 0.00}_{- 0.00}$ & $ 0.99^{+ 0.00}_{- 0.00}$ \\ 
\cline{2-12}
    & 0.95 & $ 0.95^{+ 0.00}_{- 0.00}$ & $ 0.95^{+ 0.00}_{- 0.01}$ & $ 0.94^{+ 0.01}_{- 0.01}$ & $ 0.93^{+ 0.01}_{- 0.01}$ & $ 0.91^{+ 0.01}_{- 0.02}$ & $ 0.90^{+ 0.01}_{- 0.02}$ & $ 0.93^{+ 0.02}_{- 0.01}$ & $ 0.99^{+ 0.00}_{- 0.00}$ & $ 0.99^{+ 0.00}_{- 0.00}$ & $ 0.98^{+ 0.00}_{- 0.00}$ \\ 
\cline{2-12}
\hline
\hline
\et
\ec
\etab
\egroup


\bgroup
\def\arraystretch{2.1}
\btab
\caption{Pielou index for the communities sizes identified by each algorithm for the OUR Benchmarks. The values in the table are averages over all 100 graphs for each value of $\mu$ and benchmarked (generated) pielou index.}\label{tabOURpielou}
\bc
\tiny
\bt{c||c|cccccccccc}
\multirow{2}{*}{Algorithm} & \multirow{2}{*}{Pielou} & \multicolumn{10}{c}{$\mu$} \\
 \cline{3-12}
 & & 0.0 & 0.1 & 0.2 & 0.3 & 0.4 & 0.5 & 0.6 & 0.7 & 0.8 & 0.9 \\
\hline
\hline
\multirow{3}{*}{louvain}  & 0.75 & $ 0.79^{+ 0.00}_{- 0.00}$ & $ 0.95^{+ 0.02}_{- 0.04}$ & $ 0.96^{+ 0.01}_{- 0.02}$ & $ 0.97^{+ 0.01}_{- 0.02}$ & $ 0.97^{+ 0.01}_{- 0.02}$ & $ 0.97^{+ 0.01}_{- 0.02}$ & $ 0.97^{+ 0.01}_{- 0.02}$ & $ 0.97^{+ 0.01}_{- 0.02}$ & $ 0.98^{+ 0.01}_{- 0.02}$ & $ 0.97^{+ 0.01}_{- 0.02}$ \\ 
\cline{2-12}
    & 0.85 & $ 0.90^{+ 0.01}_{- 0.00}$ & $ 0.97^{+ 0.01}_{- 0.02}$ & $ 0.98^{+ 0.01}_{- 0.01}$ & $ 0.98^{+ 0.01}_{- 0.01}$ & $ 0.98^{+ 0.01}_{- 0.01}$ & $ 0.98^{+ 0.01}_{- 0.01}$ & $ 0.97^{+ 0.01}_{- 0.02}$ & $ 0.97^{+ 0.01}_{- 0.01}$ & $ 0.98^{+ 0.01}_{- 0.02}$ & $ 0.97^{+ 0.01}_{- 0.02}$ \\ 
\cline{2-12}
    & 0.95 & $ 1.00^{+ 0.00}_{- 0.00}$ & $ 0.99^{+ 0.00}_{- 0.01}$ & $ 0.99^{+ 0.01}_{- 0.01}$ & $ 0.99^{+ 0.01}_{- 0.01}$ & $ 0.98^{+ 0.01}_{- 0.01}$ & $ 0.97^{+ 0.01}_{- 0.02}$ & $ 0.96^{+ 0.01}_{- 0.02}$ & $ 0.97^{+ 0.01}_{- 0.02}$ & $ 0.97^{+ 0.01}_{- 0.02}$ & $ 0.97^{+ 0.01}_{- 0.01}$ \\ 
\cline{2-12}
\hline
\hline
\multirow{3}{*}{CPM}  & 0.75 & $ 0.75^{+ 0.00}_{- 0.00}$ & $ 0.75^{+ 0.01}_{- 0.01}$ & $ 0.75^{+ 0.01}_{- 0.01}$ & $ 0.76^{+ 0.01}_{- 0.02}$ & $ 0.76^{+ 0.02}_{- 0.02}$ & $ 0.77^{+ 0.02}_{- 0.02}$ & $ 0.79^{+ 0.02}_{- 0.03}$ & $ 0.81^{+ 0.02}_{- 0.03}$ & $ 0.85^{+ 0.02}_{- 0.02}$ & $ 0.99^{+ 0.00}_{- 0.00}$ \\ 
\cline{2-12}
    & 0.85 & $ 0.85^{+ 0.00}_{- 0.00}$ & $ 0.85^{+ 0.00}_{- 0.01}$ & $ 0.85^{+ 0.01}_{- 0.01}$ & $ 0.85^{+ 0.01}_{- 0.01}$ & $ 0.86^{+ 0.01}_{- 0.01}$ & $ 0.86^{+ 0.01}_{- 0.02}$ & $ 0.87^{+ 0.02}_{- 0.02}$ & $ 0.88^{+ 0.02}_{- 0.02}$ & $ 0.96^{+ 0.02}_{- 0.03}$ & $ 0.99^{+ 0.00}_{- 0.00}$ \\ 
\cline{2-12}
    & 0.95 & $ 0.95^{+ 0.00}_{- 0.00}$ & $ 0.95^{+ 0.00}_{- 0.00}$ & $ 0.95^{+ 0.00}_{- 0.00}$ & $ 0.95^{+ 0.00}_{- 0.00}$ & $ 0.95^{+ 0.00}_{- 0.01}$ & $ 0.95^{+ 0.01}_{- 0.01}$ & $ 0.96^{+ 0.02}_{- 0.02}$ & $ 0.97^{+ 0.01}_{- 0.02}$ & $ 0.99^{+ 0.00}_{- 0.00}$ & $ 0.99^{+ 0.00}_{- 0.00}$ \\ 
\cline{2-12}
\hline
\hline
\multirow{3}{*}{Infomap}  & 0.75 & $ 0.79^{+ 0.00}_{- 0.00}$ & $ 0.79^{+ 0.01}_{- 0.01}$ & $ 0.81^{+ 0.02}_{- 0.02}$ & $ 0.83^{+ 0.02}_{- 0.02}$ & $ 0.85^{+ 0.02}_{- 0.03}$ & $ 0.88^{+ 0.02}_{- 0.03}$ & $ 0.91^{+ 0.03}_{- 0.03}$ & $ 0.55^{+ 0.38}_{- 0.55}$ & $ 0.00^{+ 0.14}_{- 0.00}$ & $ 0.00^{+ 0.00}_{- 0.00}$ \\ 
\cline{2-12}
    & 0.85 & $ 0.90^{+ 0.01}_{- 0.01}$ & $ 0.89^{+ 0.01}_{- 0.01}$ & $ 0.90^{+ 0.01}_{- 0.01}$ & $ 0.92^{+ 0.01}_{- 0.01}$ & $ 0.93^{+ 0.01}_{- 0.01}$ & $ 0.94^{+ 0.01}_{- 0.02}$ & $ 0.95^{+ 0.02}_{- 0.02}$ & $ 0.47^{+ 0.49}_{- 0.48}$ & $ 0.00^{+ 0.00}_{- 0.00}$ & $ 0.00^{+ 0.00}_{- 0.00}$ \\ 
\cline{2-12}
    & 0.95 & $ 1.00^{+ 0.00}_{- 0.00}$ & $ 0.98^{+ 0.01}_{- 0.01}$ & $ 0.99^{+ 0.01}_{- 0.01}$ & $ 0.99^{+ 0.01}_{- 0.01}$ & $ 1.00^{+ 0.00}_{- 0.01}$ & $ 1.00^{+ 0.00}_{- 0.00}$ & $ 0.84^{+ 0.16}_{- 0.86}$ & $ 0.13^{+ 0.88}_{- 0.13}$ & $ 0.00^{+ 0.00}_{- 0.00}$ & $ 0.00^{+ 0.00}_{- 0.00}$ \\ 
\cline{2-12}
\hline
\hline
\multirow{3}{*}{RB}  & 0.75 & $ 0.79^{+ 0.00}_{- 0.00}$ & $ 0.91^{+ 0.03}_{- 0.04}$ & $ 0.92^{+ 0.02}_{- 0.04}$ & $ 0.93^{+ 0.02}_{- 0.03}$ & $ 0.92^{+ 0.02}_{- 0.03}$ & $ 0.92^{+ 0.02}_{- 0.04}$ & $ 0.92^{+ 0.02}_{- 0.03}$ & $ 0.92^{+ 0.02}_{- 0.02}$ & $ 0.91^{+ 0.03}_{- 0.02}$ & $ 0.99^{+ 0.00}_{- 0.00}$ \\ 
\cline{2-12}
    & 0.85 & $ 0.90^{+ 0.01}_{- 0.01}$ & $ 0.94^{+ 0.01}_{- 0.02}$ & $ 0.94^{+ 0.01}_{- 0.02}$ & $ 0.94^{+ 0.01}_{- 0.02}$ & $ 0.94^{+ 0.01}_{- 0.02}$ & $ 0.94^{+ 0.01}_{- 0.02}$ & $ 0.94^{+ 0.01}_{- 0.02}$ & $ 0.93^{+ 0.02}_{- 0.01}$ & $ 0.96^{+ 0.02}_{- 0.03}$ & $ 0.99^{+ 0.00}_{- 0.00}$ \\ 
\cline{2-12}
    & 0.95 & $ 1.00^{+ 0.00}_{- 0.00}$ & $ 0.96^{+ 0.01}_{- 0.01}$ & $ 0.95^{+ 0.01}_{- 0.01}$ & $ 0.95^{+ 0.01}_{- 0.01}$ & $ 0.96^{+ 0.01}_{- 0.01}$ & $ 0.97^{+ 0.01}_{- 0.01}$ & $ 0.99^{+ 0.01}_{- 0.01}$ & $ 0.98^{+ 0.01}_{- 0.01}$ & $ 0.99^{+ 0.00}_{- 0.00}$ & $ 0.99^{+ 0.00}_{- 0.00}$ \\ 
\cline{2-12}
\hline
\hline
\multirow{3}{*}{RN}  & 0.75 & $ 0.75^{+ 0.01}_{- 0.00}$ & $ 0.69^{+ 0.07}_{- 0.69}$ & $ 0.56^{+ 0.20}_{- 0.55}$ & $ 0.52^{+ 0.25}_{- 0.50}$ & $ 0.40^{+ 0.37}_{- 0.37}$ & $ 0.30^{+ 0.48}_{- 0.27}$ & $ 0.16^{+ 0.63}_{- 0.13}$ & $ 0.06^{+ 0.18}_{- 0.03}$ & $ 0.04^{+ 0.15}_{- 0.02}$ & $ 0.09^{+ 0.81}_{- 0.06}$ \\ 
\cline{2-12}
    & 0.85 & $ 0.85^{+ 0.00}_{- 0.00}$ & $ 0.81^{+ 0.04}_{- 0.84}$ & $ 0.80^{+ 0.06}_{- 0.82}$ & $ 0.73^{+ 0.12}_{- 0.73}$ & $ 0.62^{+ 0.25}_{- 0.60}$ & $ 0.45^{+ 0.42}_{- 0.42}$ & $ 0.24^{+ 0.64}_{- 0.21}$ & $ 0.09^{+ 0.70}_{- 0.06}$ & $ 0.03^{+ 0.03}_{- 0.01}$ & $ 0.07^{+ 0.51}_{- 0.05}$ \\ 
\cline{2-12}
    & 0.95 & $ 0.95^{+ 0.00}_{- 0.00}$ & $ 0.95^{+ 0.00}_{- 0.00}$ & $ 0.95^{+ 0.00}_{- 0.00}$ & $ 0.94^{+ 0.01}_{- 0.11}$ & $ 0.94^{+ 0.01}_{- 0.10}$ & $ 0.91^{+ 0.05}_{- 0.94}$ & $ 0.75^{+ 0.22}_{- 0.74}$ & $ 0.19^{+ 0.80}_{- 0.17}$ & $ 0.10^{+ 0.90}_{- 0.07}$ & $ 0.08^{+ 0.75}_{- 0.05}$ \\ 
\cline{2-12}
\hline
\hline
\multirow{3}{*}{SCluster}  & 0.75 & $ 0.76^{+ 0.00}_{- 0.00}$ & $ 0.75^{+ 0.01}_{- 0.02}$ & $ 0.75^{+ 0.01}_{- 0.01}$ & $ 0.76^{+ 0.01}_{- 0.02}$ & $ 0.76^{+ 0.02}_{- 0.02}$ & $ 0.77^{+ 0.02}_{- 0.02}$ & $ 0.78^{+ 0.02}_{- 0.02}$ & $ 0.78^{+ 0.02}_{- 0.02}$ & $ 0.85^{+ 0.02}_{- 0.02}$ & $ 0.99^{+ 0.00}_{- 0.00}$ \\ 
\cline{2-12}
    & 0.85 & $ 0.86^{+ 0.00}_{- 0.00}$ & $ 0.85^{+ 0.00}_{- 0.01}$ & $ 0.85^{+ 0.01}_{- 0.02}$ & $ 0.85^{+ 0.01}_{- 0.01}$ & $ 0.86^{+ 0.01}_{- 0.01}$ & $ 0.86^{+ 0.01}_{- 0.02}$ & $ 0.85^{+ 0.02}_{- 0.01}$ & $ 0.86^{+ 0.01}_{- 0.01}$ & $ 0.96^{+ 0.01}_{- 0.02}$ & $ 0.99^{+ 0.00}_{- 0.00}$ \\ 
\cline{2-12}
    & 0.95 & $ 0.95^{+ 0.01}_{- 0.01}$ & $ 0.95^{+ 0.01}_{- 0.01}$ & $ 0.95^{+ 0.00}_{- 0.00}$ & $ 0.95^{+ 0.00}_{- 0.00}$ & $ 0.95^{+ 0.01}_{- 0.01}$ & $ 0.95^{+ 0.01}_{- 0.01}$ & $ 0.94^{+ 0.01}_{- 0.01}$ & $ 0.95^{+ 0.01}_{- 0.01}$ & $ 0.99^{+ 0.00}_{- 0.00}$ & $ 0.99^{+ 0.00}_{- 0.00}$ \\ 
\cline{2-12}
\hline
\hline
\multirow{3}{*}{UVCluster}  & 0.75 & $ 0.76^{+ 0.00}_{- 0.00}$ & $ 0.76^{+ 0.01}_{- 0.01}$ & $ 0.77^{+ 0.01}_{- 0.01}$ & $ 0.79^{+ 0.02}_{- 0.02}$ & $ 0.80^{+ 0.02}_{- 0.02}$ & $ 0.82^{+ 0.02}_{- 0.02}$ & $ 0.83^{+ 0.02}_{- 0.02}$ & $ 0.84^{+ 0.02}_{- 0.02}$ & $ 0.97^{+ 0.02}_{- 0.05}$ & $ 0.99^{+ 0.00}_{- 0.00}$ \\ 
\cline{2-12}
    & 0.85 & $ 0.86^{+ 0.00}_{- 0.00}$ & $ 0.86^{+ 0.01}_{- 0.01}$ & $ 0.86^{+ 0.01}_{- 0.01}$ & $ 0.87^{+ 0.01}_{- 0.01}$ & $ 0.88^{+ 0.01}_{- 0.01}$ & $ 0.88^{+ 0.01}_{- 0.02}$ & $ 0.89^{+ 0.01}_{- 0.01}$ & $ 0.90^{+ 0.01}_{- 0.01}$ & $ 0.99^{+ 0.00}_{- 0.00}$ & $ 0.99^{+ 0.00}_{- 0.00}$ \\ 
\cline{2-12}
    & 0.95 & $ 0.95^{+ 0.01}_{- 0.01}$ & $ 0.95^{+ 0.01}_{- 0.01}$ & $ 0.96^{+ 0.01}_{- 0.01}$ & $ 0.97^{+ 0.01}_{- 0.01}$ & $ 0.97^{+ 0.01}_{- 0.01}$ & $ 0.97^{+ 0.01}_{- 0.01}$ & $ 0.96^{+ 0.01}_{- 0.01}$ & $ 0.98^{+ 0.01}_{- 0.01}$ & $ 0.99^{+ 0.00}_{- 0.00}$ & $ 0.99^{+ 0.00}_{- 0.00}$ \\ 
\cline{2-12}
\hline
\hline
\multirow{3}{*}{surpriser}  & 0.75 & $ 0.75^{+ 0.00}_{- 0.00}$ & $ 0.75^{+ 0.01}_{- 0.01}$ & $ 0.76^{+ 0.01}_{- 0.01}$ & $ 0.76^{+ 0.01}_{- 0.02}$ & $ 0.76^{+ 0.02}_{- 0.02}$ & $ 0.77^{+ 0.02}_{- 0.02}$ & $ 0.79^{+ 0.02}_{- 0.03}$ & $ 0.80^{+ 0.02}_{- 0.03}$ & $ 0.92^{+ 0.05}_{- 0.06}$ & $ 0.99^{+ 0.00}_{- 0.00}$ \\ 
\cline{2-12}
    & 0.85 & $ 0.85^{+ 0.00}_{- 0.00}$ & $ 0.85^{+ 0.01}_{- 0.01}$ & $ 0.85^{+ 0.01}_{- 0.01}$ & $ 0.85^{+ 0.01}_{- 0.01}$ & $ 0.85^{+ 0.01}_{- 0.01}$ & $ 0.86^{+ 0.01}_{- 0.02}$ & $ 0.87^{+ 0.02}_{- 0.02}$ & $ 0.88^{+ 0.02}_{- 0.02}$ & $ 0.99^{+ 0.00}_{- 0.01}$ & $ 0.99^{+ 0.00}_{- 0.00}$ \\ 
\cline{2-12}
    & 0.95 & $ 0.95^{+ 0.00}_{- 0.00}$ & $ 0.95^{+ 0.00}_{- 0.01}$ & $ 0.95^{+ 0.00}_{- 0.01}$ & $ 0.95^{+ 0.01}_{- 0.01}$ & $ 0.95^{+ 0.00}_{- 0.01}$ & $ 0.95^{+ 0.01}_{- 0.01}$ & $ 0.95^{+ 0.01}_{- 0.02}$ & $ 0.96^{+ 0.01}_{- 0.02}$ & $ 0.99^{+ 0.00}_{- 0.00}$ & $ 0.99^{+ 0.00}_{- 0.00}$ \\ 
\cline{2-12}
\hline
\hline
\et
\ec
\etab
\egroup

\subsection{A Survey of the Surprise Landscape} \label{secLandscape}

The number of different partitions into which one may divide the nodes of a network, even for graphs with only a few nodes, is astronomically huge and so it is not possible to test them all in search for the one that results in the global maximum of a quality function, therefore the need for heuristic algorithms. Such algorithms may have drawbacks like returning a local and not the global maximum, which in any case may be degenerated. An important question to tackle is whether having two partitions laying at two different local maxima with close values, mean that they represent the system in similar configurations. If they do, the problem of whether the algorithm is returning the global or a nearby local maximum is diminished for, even though the algorithm may return a local maximum, the closer it is to the global, the more the solution represent the system in the optimal configuration. 

The goal here is, therefore, to analyse the difference between the partitions in an abstract partition space and the relationship between their values of Modularity and its competitor, the Surprise. For comparing different partitions we will be using the VI, a properly defined metric. In particular, given many ($N$) different partitions for the nodes in a network, we start by creating a distance matrix $d_{ij}$, where every element is the VI between two partitions $i$ and $j$. In order to make an intuitive visualization of this matrix, we follow a similar approach as done in \cite{linearembedding}. We proceed to evaluate an embedding of all the partitions: a vector $\vec{r}_i=(x_i,y_i)$ is associated to each partition in such a way that the euclidean distance between two different vectors $\vec{r}_i$ and $\vec{r}_j$, each representing a different partition, is as close as possible to the VI between partitions $i$ and $j$.

This embedding is created using a steepest descent algorithm that walks over the $\vec{r}_i$ ($2N$-dimensional) parameter space searching for the minimum of $\chi^2$ defined as:

\be
\chi^2 &=& \sum_{i=1}^N \sum_{j>i|d_{ij}<d_{\textrm{lim}}} d_{ij}^{\gamma}\left(d_{ij} - \sqrt{(x_i-x_j)^2 + (y_i-y_j)^2}\right)^2, \label{chisquare}
\ee
where the first sum runs over all partitions while the second sum runs over the partitions whose distance to the $i$'th partition is less than some predetermined value $d_{\textrm{lim}}$. The factor $d_{ij}^{\gamma}$ allows to use the parameter $\gamma$ in order to control the relative importance of partitions closer or further apart\footnote{No relation of this $\gamma$ to the parameters used in order to generate benchmarks.}. The smaller the value of $\chi^2$, the closer the euclidean distances are to the VI between partitions $i$ and $j$ ($d_{ij}$). To minimize $\chi^2$ we evaluate the gradient of $\chi^2$ in parameter space and then update the values of $x_i$ and $y_i$ in the direction opposite of this gradient, so going towards smaller values of $\chi^2$.

In order to thoroughly study the partition space of a network, we choose to work with a simple graph: the one studied in \cite{aldecoa2010} and shown in figure \ref{simplenet}. This 11 nodes graph is composed by two size 4 cliques connected by a length 3 path. In the partition with maximum modularity (global maximum), each 4-clique is assigned to a different community and the 3-path to a third community. In the case of the surprise, the global maximum is degenerated: each 4-clique forms a separated community, but one of the extreme nodes in the 3-path is assigned to a community alone and the other two nodes to a fourth community and, since there are two way to make this partition, there are two different partitions with the same maximum surprise value. 

\bfig
\bc
\ig{0.6}{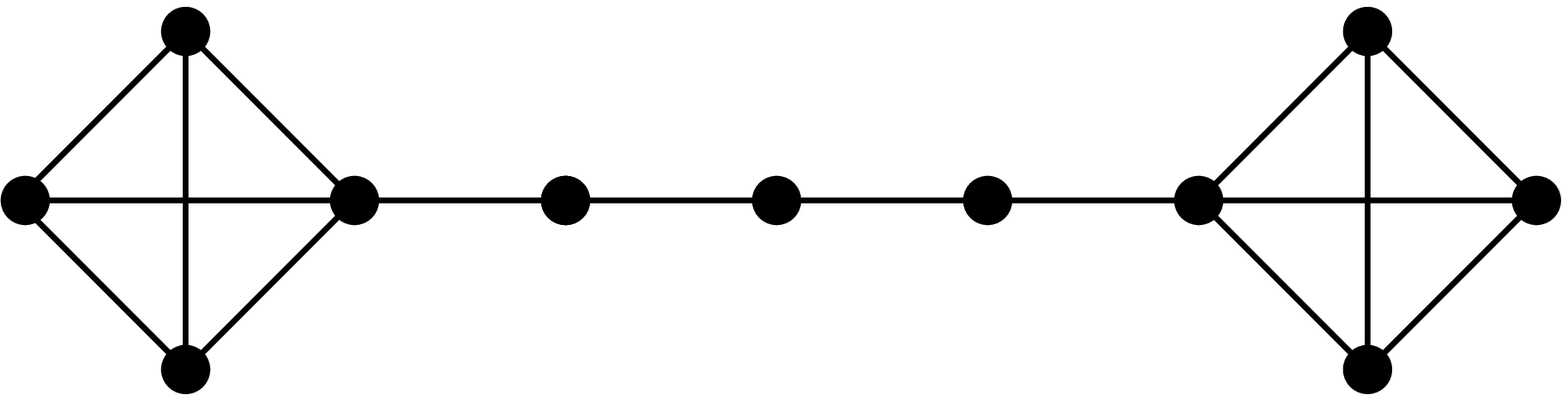}
\ec
\caption{Graph for which the surprise and modularity surface over partition space was studied.} \label{simplenet}
\efig

For this network, we produced 1336 different partitions by means of the annealing functions implemented in the surpriser package. The embedding was produced using as parameters $\gamma=-1$ (giving more importance to partitions close by) and $d_{\textrm{lim}}=1.$ (considering all distances between every pair of partitions). Figures \ref{fig:surface} a) and b) show the resulting modularity and surprise surfaces resulting from the embedding: over the xy-plane are the resulting coordinates representing each partition and in the z-axis the value of Modularity and Surprise respectively.

In figure \ref{fig:surface} one can compare the surprise and modularities landscapes for the set of 1336 partitions of the graph in figure \ref{simplenet}. One sees that in Modularity's case, although the global maximum is not degenerated, the degeneracy problem is accentuated, as there is a plateau region where many different solutions lie really close to the global maximum in the z-axis, but are far apart in the xy-plane. Because the distance in the xy-plane mimics the VI between two partitions, this plateau region contains partitions that are close to the global maximum in value but that are very different from each other in structure (represent very different network partitions). Surprise, on the other hand, has two sharp peaks that are the optimal value, close to a third representing the partition that results in the modularity maximum, all close together. As the points near the maximum are closer to each other, this means that the close to optimal partitions are also more similar and the further away one is to the region of the maxima, the more different the partitions represented are and the smaller and smaller the values in surprise.

This does not mean that Surprise does not suffer from a degeneracy problem at all,  what it does mean is that different partitions with a corresponding surprise value close to the maximum are similar to each other, such that even if an algorithm doesn't find the global maximum, the partitions that are nearby that point are representative of the network's true community structure. Both Surprise and Modularity suffer from the degeneracy problem in the sense that there's no guarantee that the global maximum is unique, however partitions close to the maximum of Surprise are similar to each other, unlike Modularity, where two partition with high values of modularity may be representing different community divisions \cite{good2010performance}.

\bfig
\bc
a) Modularity

\includegraphics[width=0.70\textwidth]{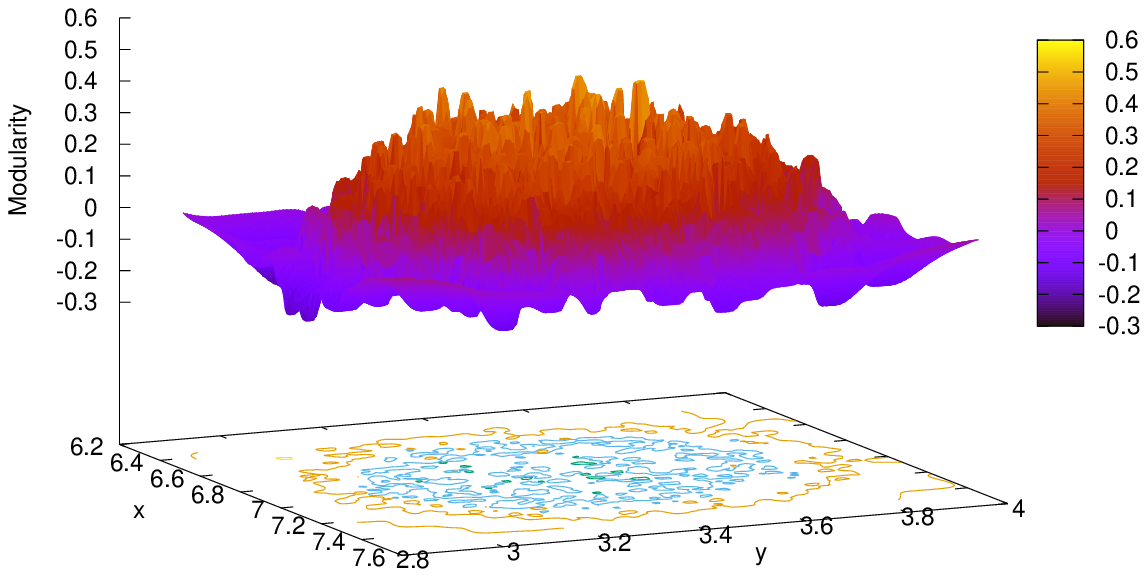}

\hrule{}

\ssp

b) Surprise

\includegraphics[width=0.70\textwidth]{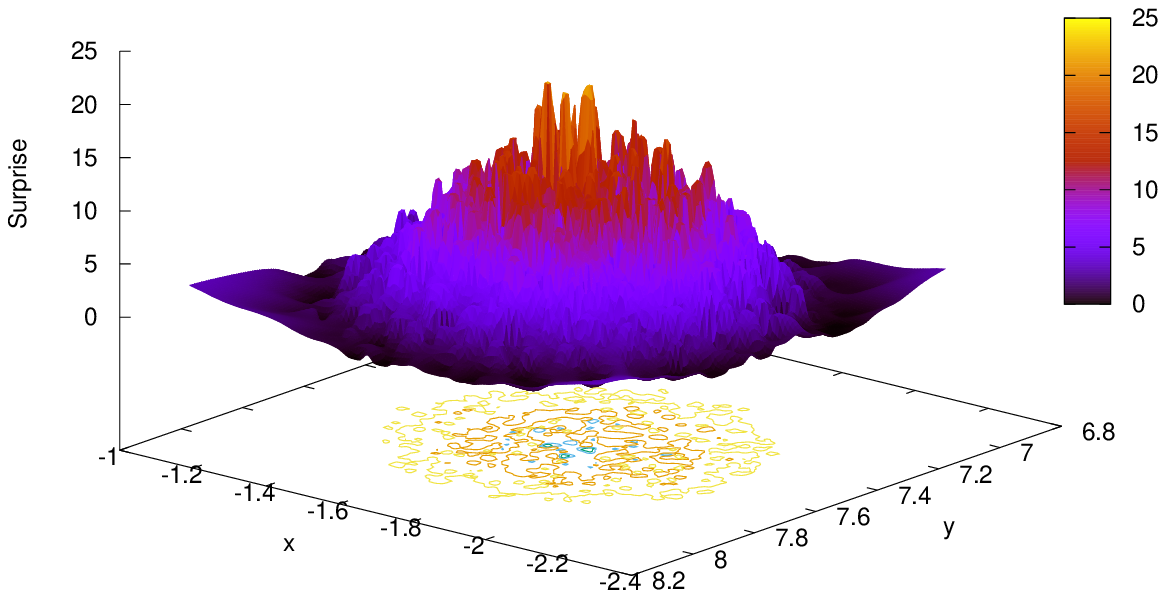}
\caption{Surprise and modularity landscapes over partition space.}
\label{fig:surface}
\ec
\efig

In order to stress this point further and more objectively, we also produced the plot in figure \ref{figdist}. In this plot, since, the two functions have different bounds, we first normalize each surface such that the highest peak has height 1 and the lowest point is at zero. Next we sort the all partitions by height and imagine a walk starting from the highest peak to the next highest one and to the next, and so one until the hundredth one. In the x-axis of the plot is the cumulative distance, the sum of the VIs between two consecutive partitions until the point one is at and in the y-axis the height of this point. One clearly sees from the figure that the surprise highest points are much more prominent than the ones in the modularity surface and, moreover, one walks a greater distance in the modularity surface in order to cover the same number of points, indicating that they are, on average, further a part from each other. We note however, that the graph over which these partitions were made is a rather uninvolved one and, understanding the shortcuts of modularity, it is clear that having a more complex graph would only exacerbate the issues pointed out by this analysis. 

\bfig
\bc
\ig{0.6}{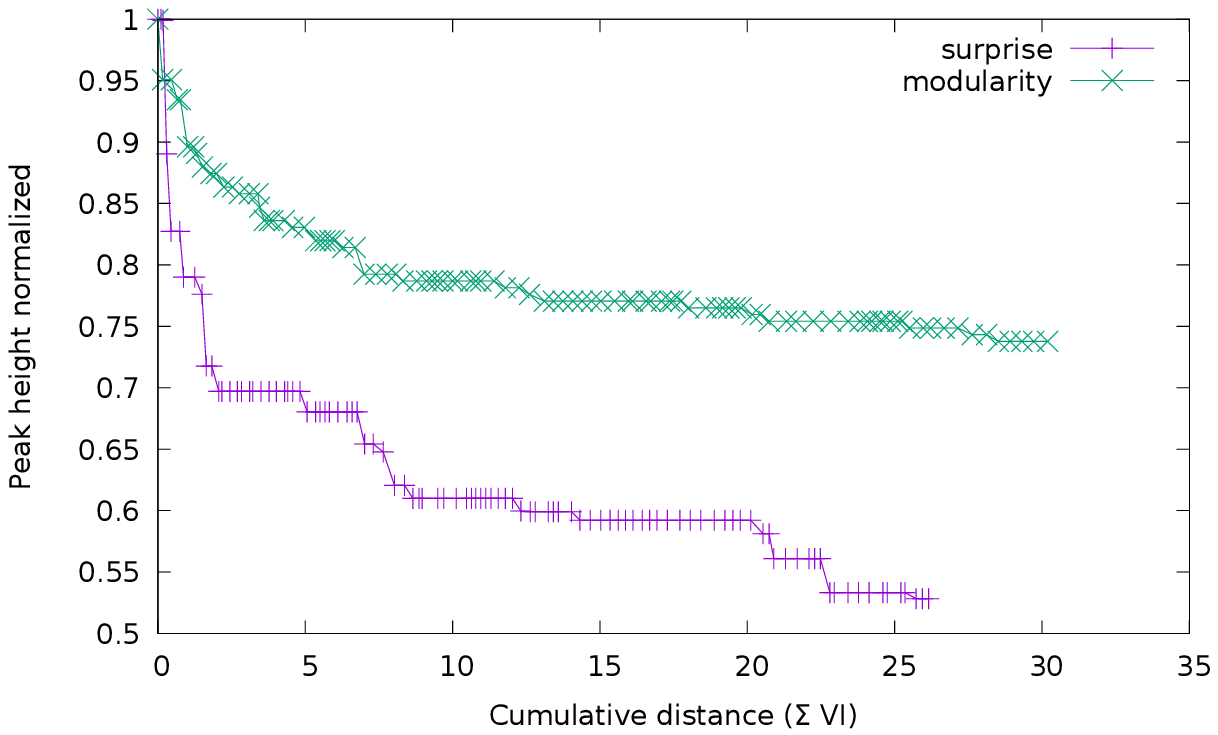}
\ec
\caption{Walk from the highest peak in each surface to every other.} \label{figdist}
\efig


\section{Overview and Conclusions}

The definition of communities in a graph and the best way to detect such structures are open problems in graph theory. As a consequence, several different approaches to identify these groups have been suggested, as well as different validation benchmarks for these different algorithms. Therefore it becomes important a systematical and objective comparison of these methods.

In the present work, we started by generating a number of benchmark networks following two recipes normally employed for this objective (LFR and RC) and a third one that we propose in order to isolate and control separately the different issues that these benchmarks may present. Next we compare the results obtained from different community detection algorithms over these graphs and search for the possible biases of each one over each benchmark. We point out that the networks generated by the LFR benchmark may suffer from the fact that they arbitrarily assign nodes to communities fulfilling requirements at a local (node) level and not at global (community) level. As a result, the communities formed may present subsets of vertices that are poorly connected to the rest of the community and algorithms that do not suffer from the issue known as resolution problem tend to break these communities into smaller pieces. One observes therefore that algorithms that perform well in the LFR benchmark tend to perform worse in the other two where the communities are defined from cliques which are slowly degraded. It is not for us to say which is the correct way to define a community, but those applying these techniques in a particular system with particular goals, should be aware of these biases and issues in order to correctly select the algorithm or the community definition more in line with their objectives and systems under study. Such algorithms with a bad resolution also tend to divide the graphs into communities of similar sizes and therefore have problems identifying communities of heterogeneous sizes. 

We also presented a new algorithm based on the optimization of the quality function named as surprise. We showed that this method is able to resolve communities of heterogeneous sizes, resolving big and small communities in the same graph, as opposite to algorithms that tend to group nodes in more homogeneous clusters, like louvain. Moreover we showed that this quality function has a steeper maximum region, meaning that the higher the surprise value for a given partition, the closer this partition is to the optimal one. This is in contrast to the modularity landscape, where very different partitions may have very similar values of modularity close to the global maximum.


\section*{Acknowledgments}

J.~A.~Pellizaro would like to thank financial support from the {\it Coordenação de Aperfeiçoamento de Pessoal de Nível Superior} – Brasil (CAPES).


\appendix

\section{The Surpriser Package} \label{appendixPython}

The Surpriser package can be found in the The Python Package Index (PyPI - https://pypi.org/project/Surpriser/). Along with the source code and installation files, one finds also a comprehensive manual where each function and method is thoroughly explained with details on how to use the functions arguments in order to perform in the best way each possible operation. In the manual, each explanation is illustrated with an useful example, some of them reproducing a few results found in this article. Below we briefly highlight the functionalities and the data that can be found in the package and after that we comment on some numerical issues about the implementation.

\subsection{Package modules, objects and functions}

The package has four modules. Each module has the following objects and functions:

\bi
\item Module \texttt{surprise}:
    \bi
    \item \texttt{Surpriser} (Object) \arr gathers the graph structure information and has methods to work out the partition that optimizes the surprise.
    \item \texttt{compare} (Function) \arr evaluates the non-normalized variation of information between two partitions (number between 0 and $\ln K$, where $K$ is the length of the lists being compared).
    \item \texttt{surprise} (Function) \arr evaluates the value for the function surprise found in equation (\ref{surprise}).
    \item \texttt{fact} (Function) \arr returns the natural logarithm of the factorial of a number ($\ln n!$).
    \item \texttt{gammas} (Function) \arr returns the natural logarithm of a combination ($\ln \left(\ba{c} n \\ k \ea \right)=\ln n! - \ln k! - ln(n-k)!$).
    \item \texttt{ChiGrad} (Function) \arr given the coordinates of an embedding and a matrix of distances $d_{ij}$, evaluates the $\chi^2$ in equation (\ref{chisquare}), its gradient in parameter space and the modulus of this gradient.
    \item \texttt{embedding} (Function) \arr given a matrix of distances $d_{ij}$, evaluates the coordinates $\vec{r}_i=(x_i, y_i)$ that minimize $\chi^2$ in equation (\ref{chisquare}).
    \ei
\item Module \texttt{data}:
    \bi
    \item \texttt{Mtoy} (List of lists of integers) \arr M matrix for the graph in figure \ref{simplenet}.
    \item \texttt{Msyn} (List of lists of integers) \arr M matrix for the metabolic network of {\it synechocystis sp. PCC6803}. This network was constructed using information from the KEGG database \cite{kegg} as explained in \cite{gamermann2}.
    \item \texttt{NODESsyn} (List of strings) \arr Metabolite represented by each node in the Msyn network.
    \item \texttt{M243273} (List of lists of integers) \arr M matrix for the PPI network of {\it Mycoplasma genitalium}. This network was constructed using information from the STRING database \cite{string} as explained in \cite{gamermann2}.    
    \item \texttt{NODES243273} (List of strings) \arr Protein represented by each node in the M243273 network.
    \item \texttt{partsToy} (List of lists of integers) \arr List of 1336 partitions of the graph in figure \ref{simplenet}.
    \item \texttt{surpsToy} (List of floats) \arr Surprise value of the partitions in \texttt{partsToy}.
    \item \texttt{comsToPart} (Function) \arr Transforms a list of communities into a partition.
    \item \texttt{partToComs} (Function) \arr Transforms a partition into a list of communities.
    \item \texttt{surStats} (Function) \arr Given a \texttt{Surpriser} instance current state, evaluates the parameters $p$, $q$ and $r$ and other information from its community structure.
    \item \texttt{check} (Function) \arr Checks the integrity of an instance of the Surpriser object. This was programmed for debugging purposes. 
    \ei
\item Module \texttt{randoms}:
    \bi
    \item \texttt{drand} (Function) \arr Returns a random number with homogeneus distribution between 0 and 1.
    \item \texttt{drand\_SF} (Function) \arr Returns a random number with continuous scale-free distribution. 
    \item \texttt{irand\_SF} (Function) \arr Returns a random number with discrete scale-free distribution.
    \item \texttt{gamma\_MLE\_cont} (Function) \arr Given a list of floats, evaluates by maximum likelihood the best $\gamma$ supposing the numbers come from a continuous scale-free distribution ($p(k)\propto k^{-\gamma}$, $k\in\mathds{R}^+$).
    \item \texttt{gamma\_MLE} (Function) \arr Given a list of integers, evaluates by maximum likelihood the best $\gamma$ supposing the numbers come from a discrete scale-free distribution ($p(k)\propto k^{-\gamma}$, $k\in\mathds{N}$). Details on the equation one needs to solve in order to obtain $\gamma$ can be found in \cite{gamermann2}.
    \item \texttt{lnL\_cont} (Function) \arr Given a list of floats, evaluates the log-likelihood supposing the numbers were generated by a continuous scale-free distribution.
    \item \texttt{lnL} (Function) \arr Given a list of integers, evaluates the log-likelihood supposing the numbers were generated by a discrete scale-free distribution.
    \item \texttt{zeta} (Function) \arr Riemann zeta function in equation (\ref{riemannzeta}). The sum in this equation converges very slowly, but in the implementation of this function we used the method developed in \cite{accelerating}. 
    \item \texttt{dzetadx} (Function) \arr First derivative of the Riemann zeta function in equation (\ref{riemannzeta}) with respect to $\gamma$.
    \item \texttt{stats} (Functions) \arr Given a list of numbers evaluates the list's average, standard deviation and (optionally) the skewness.
    \ei
\item Module \texttt{benchmark}:
    \bi
    \item \texttt{MBenchmark} (Object) \arr Instance of a benchmark network of type OUR that can be degraded via $p$, and $q$. The initial structure is constructed from a list of numbers (cliques sizes), the parameter $r$ and an optional argument indicating whether to connect the cliques in a ring.
    \item \texttt{Pielou} (Function) \arr Given a list of numbers representing community sizes, evaluates the corresponding pielou index.
    \item \texttt{Pielouer} (Function) \arr Creates a list of numbers with a given controlled pielou index.
    \item \texttt{PielouerNodes} (Function) \arr Creates a list of numbers with a given controlled pielou index with the possibility of also fixing within a given range the total sum of the numbers (total number of nodes in the clique set).
    \ei
\ei

For the two objects in the modules (\texttt{Surpriser} and \texttt{MBenchmark}), we list below their attributes and methods (functions). These objects have attributes that are fixed and do not change and others that depend on their current states. The state of each object may change according to the methods (functions) called. The \texttt{Surpriser} may have its current partition changed by its methods, while \texttt{MBenchmark} may have the links inside the network created or destroyed by its methods.

\bi
\item \texttt{Surpriser} object:
    \bi
    \item \texttt{M} (Integer) \arr Total number of possible links inside communities, given the current network's partition.
    \item \texttt{F} (Integer) \arr Total number of possible links in the network, given its size.
    \item \texttt{K} (Integer) \arr Number of nodes in the network.
    \item \texttt{nl} (Integer) \arr Total number of links in the network.
    \item \texttt{p} (Integer) \arr Number of links between nodes in a same community, given the current partition of the graph into communities.
    \item \texttt{Nc} (Integer) \arr Number of different communities in the current partition.
    \item \texttt{surprise} (Float) \arr Surprise value corresponding to the current partition.
    \item \texttt{connected} (Function) \arr Checks whether two nodes are connected.
    \item \texttt{partition} (Function) \arr Returns the current partition of the network.
    \item \texttt{community} (Function) \arr Returns a given community from the current partition of the network.
    \item \texttt{show\_communities} (Function) \arr Prints on screent the current division of the graph into communities.
    \item \texttt{linksin} (Function) \arr Given a list of nodes, returns a tuple indicating the total number of links between the nodes and the number of links to nodes outside the list.
    \item \texttt{extractor} (Function) \arr Tries to extract one or many nodes from communities.
    \item \texttt{extractor\_an} (Function) \arr Same as \texttt{extractor}, but the operation is performed as an annealing with a given temperature.
    \item \texttt{exchanger} (Function) \arr Tries to exchange elements between communities.
    \item \texttt{exchanger\_an} (Function) \arr Same as \texttt{exchanger}, but the operation is performed as an annealing with a given temperature.
    \item \texttt{merger} (Function) \arr Tries to merge pairs of communities into a single one.
    \item \texttt{merger\_an} (Function) \arr Same as \texttt{merger}, but the operation is performed as an annealing with a given temperature.
    \item \texttt{subcommuniter} (Function) \arr Tries to extract a subcommunity from a community.
    \item \texttt{subcommuniter\_an} (Function) \arr Same as \texttt{subcommuniter}, but the operation is performed as an annealing with a given temperature.
    \item \texttt{subcommunity\_exchanger} (Function) \arr Tries to exchange a subcommunity between communities.
    \item \texttt{subcommunity\_exchanger\_an} (Function) \arr Same as \texttt{subcommunity\_exchanger}, but the operation is performed as an annealing with a given temperature.
    \item \texttt{subcommunity} (Function) \arr The method returns an instance of the \texttt{Surpriser} object built with the sub-graph formed by a given community.
    \item \texttt{stepper} (Function) \arr Performs all operations sistematically to exhaustion until the surprise value can no longer be raised. The method returns a tuple indicating the number of successful different operations made.
    \item \texttt{montecarlo\_step} (Function) \arr Performs a full monte carlo step over the annealing operations.  The method returns the number of successful different operations made.
    \item \texttt{checkN} (Function) \arr Returns a list with the change value in surprise that would result by all possible operations (but without actually performing the operations).
    \item \texttt{shake} (Function) \arr Performs all exchanges and subcommunity exchanges that result in a partition with the exact same value for the surprise (degenerate state).  The method returns a tuple indicating the number of exchanges and subcommunity exchanges made.
    \ei
\item \texttt{MBenchmark} object:
    \bi
    \item \texttt{M} (List of lists of integers) \arr Graph's M matrix.
    \item \texttt{K} (Integer) \arr Number of nodes in the graph. 
    \item \texttt{r} (Float) \arr Fraction of network nodes alone in their own communities in the benchmarked partition.
    \item \texttt{Nc} (Integer) \arr Number of communities in the benchmarked partition.
    \item \texttt{N1} (Integer) \arr Number of nodes in their own communities in the benchmarked partition.
    \item \texttt{Ncliques} (Integer) \arr Number of cliques used to construct the benchmark network.
    \item \texttt{NinCliques} (Integer) \arr Number of nodes inside cliques.
    \item \texttt{cliques} (List of integers) \arr List with the cliques sizes used to construct the benchmark network.
    \item \texttt{partition} (List of integers) \arr Benchmarked partition. 
    \item \texttt{cycle} (Boolean) \arr Indicates whether the network was constructed as a ring of cliques.
    \item \texttt{betweenc} (Integer) \arr Current number of links between nodes of different communities.
    \item \texttt{incliques} (Integer) \arr Current number of connections between nodes in the same communities.
    \item \texttt{func1} (lambda Function) \arr Function to control the degradation of links inside cliques.
    \item \texttt{params1} (Tuple) \arr Parameters used in \texttt{func1}.
    \item \texttt{func2} (lambda Function) \arr  Function to control the arising of links between different communities.
    \item \texttt{params2} (Tuple) \arr  Parameters used in \texttt{func2}.
    \item \texttt{degradP} (Function) \arr Destroy links inside communities according to the probabilities controlled by \texttt{func1} and \texttt{params1}.
    \item \texttt{degradQ} (Function) \arr  Create links between different communities according to the probabilities controlled by \texttt{func2} and \texttt{params2}.
    \item \texttt{saveEdges} (Function) \arr Saves to a file the current list of links in the network.
    \item \texttt{savePartition} (Function) \arr Saves to a file the benchmarked partition of the network.
    \ei
\ei

\subsection{Evaluation of the surprise}

For real graphs with a reasonable partition, the surprise in equation (\ref{surprise}) is a big number, which means that the elements in the sum are very small, sometimes even smaller than computer precision is able to deal with. For example, the value of surprise reached by the algorithm for the metabolic network of {\it synechocystis sp. PCC6803} is 

\bc $S=3748.366873$, \ec

\begin{flushleft}
which means that the sum in the right hand side of equation (\ref{surprise}) is 
\end{flushleft}

\bc $\exp(-3748.366873)$ \ec

\begin{flushleft}
and any computer would evaluate it as being zero (below machine precision). 
\end{flushleft}

In order to avoid such problems in the evaluation of $S$, the implemented function first, instead of evaluating the combination ratios in the sum (which are themselves even smaller numbers), evaluates their logarithms\footnote{Here one also understand why the functions \texttt{fact} and \texttt{gammas} were programmed in the \texttt{surprise} module in the first place.}, it then identifies the maximum value found (the one with the less negative logarithm) and extracts it from the sum:

\be
a_j &=& \frac{\binom{M}{j+\ell} \binom{F-M}{n-j-\ell}}{\binom{F}{n}} \\
S &=& -\ln\Sum_{j=0}^{\textrm{min($M-\ell$,$n-\ell$)}} a_j \nn\\
  &=& -\ln\left( a^{max }\left(1+\Sum_{j=1}^{\textrm{min($M-\ell$,$n-\ell$)}} \frac{a_j}{a^{max}}\right)\right) \nn\\
   &=& -\ln a^{max } - \ln\left(1+\Sum_{j=1}^{\textrm{min($M-\ell$,$n-\ell$)}} \frac{a_j}{a^{max}}\right), \label{newsurp}
\ee
supposing the first term in the sum was the biggest $a$ ($a^{max}=a_0$).

Now the surprise in equation (\ref{newsurp}) can be easily evaluated by a computer for the term $\ln a^{max }$, which is the dominating one, is just a short sum of logarithms of factorials and the second term is just a correction to it which can also be evaluated since the $a$ terms are all of similar order of magnitude such that one does not run into deep computer precision problems in its evaluation. We should note here that in the implementation of the SurpriseMe tools \cite{aldecoa2013} the authors have used Stirling's approximation in order to compute the value of $S$. In our implementation the result obtained is not an approximation and if the two are compared a small difference can be noticed especially for smaller values of $S$, when the approximation is worse.

\subsection{The embedding}

The embedding proposed in section \ref{secLandscape} (obtaining the coordinates $\vec{r}_i$ that minimize equation (\ref{chisquare})) is achieved through a gradient descent algorithm. The gradient of this function is a $2N$ dimensional vector, where $N$ is the number of points for which one wishes to find the coordinates. In the case described in this paper, $N=1336$.

In using this function one should be aware of numerical instabilities. The programmed function has two parameters that control how greedily the algorithm descents toward the minimum: \texttt{lamb} ($\lambda$) with default value of 0.2 which tells how fast initially the algorithm walks in the direction of the gradient ($\lambda \vec{\nabla}_{2N}\chi^2$) and \texttt{adj} with default value of 0.05 which either increases the value of $\lambda$ by a fraction proportional to it in a successfull step in the gradient direction or reduces it every non-successfull step. 

The gradient descent stops running and returns the current coordinate values in case the module of the gradient divided by the number of parameters ($|\vec{\nabla}_{2N}\chi^2|/(2N)$) is smaller than an adjustable precision (whose default value is $\epsilon=10^{-10}$) or the value of $\lambda$ goes below $10^{-18}$ or the algorithm fails to advance in the gradient direction for more than 5000 consecutive steps.

After running the embedding function, one can always obtain the $\chi^2$ value, the gradient of it in coordinate space and its module in order to check the quality of the obtained embedding by using the \texttt{ChiGrad} function in the \texttt{surprise} module.

\section{The Pielou Index} \label{appendixPielou}

The Pielou index measures how even a partition of elements into communities is. It is basically the entropy for the distribution given by the fraction of nodes in each community.

Let's take a number $K$ of nodes in a graph and divide them into $N$ communities. Each group $i$ has $c_i$ elements in it, therefore,

\bc $\Sum_{i=1}^{N}c_i=K$. \ec

Now, take a node at random. The probability that it belongs to community $i$ is $p_i=\frac{c_i}{K}$. The more even the communities sizes are, the more uncertain one can be on the community of a randomly selected element. The entropy for this distribution is a measure of this uncertainty:

\be
H &=& -\Sum_{i=1}^{N}p_i\ln p_i,
\ee
its maximum value is $\ln N$ in the case that $p_i=\frac{1}{N}$ (most even case) and its minimum is 0 if one has all elements in the same community. Thus, one can construct a normalized index (between 0 and 1) dividing $H$ by $\ln N$ and this is precisely the Pielou index:

\be
PI &=& \frac{H}{\ln N}.
\ee


\bibliographystyle{unsrt}
\bibliography{surpriser}

\end{document}